\documentclass[aps,pre,twocolumn,superscriptaddress]{revtex4}

\usepackage{epsfig,amsmath}
\usepackage{graphicx}
\usepackage{dcolumn}
\usepackage{stmaryrd}
\usepackage{mathrsfs}
\usepackage{pifont}
\usepackage{amsthm}
\usepackage{amssymb}
\usepackage{bm}
\usepackage{latexsym}
\usepackage[colorlinks=true,linkcolor=blue,citecolor=blue]{hyperref}
\usepackage{color}
\usepackage[capitalise]{cleveref}
\theoremstyle{plain}

\usepackage{subcaption}
\DeclareCaptionJustification{justified}{\leftskip=0pt \rightskip=0pt \parfillskip=0pt plus 1fil}

\usepackage{mathtools}

\begin{document}

\title{Chaos in the quantum Duffing oscillator in the semiclassical regime under parametrized dissipation}

\author{Andrew D. Maris}
\thanks{These two authors contributed equally to this work.}
\affiliation{Department of Physics and Astronomy, Carleton College, One North College Street,
Northfield, MN 55057}
\affiliation{Department of Nuclear Science and Engineering, Massachusetts Institute of Technology, 77 Massachusetts Ave, Cambridge, MA 02139}
\author{Bibek Pokharel}
\thanks{These two authors contributed equally to this work.\\ Corresponding Author: \hyperlink{bbk.pokharel@gmail.com}{bbk.pokharel@gmail.com}}
\affiliation{Department of Physics, University of Southern California, Los Angeles, CA 90089}
\author{Sharan Ganjam Seshachallam}
\author{Moses Z.R. Misplon}
\author{Arjendu K. Pattanayak}
\affiliation{Department of Physics and Astronomy, Carleton College, One North College Street,
Northfield, MN 55057}


\begin{abstract}

  We study the quantum dissipative Duffing oscillator across a range of system sizes and environmental couplings under varying semiclassical approximations. Using spatial (based on Kullback-Leibler distances between phase-space attractors) and temporal (Lyapunov exponent-based) complexity metrics, we isolate the effect of the environment on quantum-classical differences. Moreover, we quantify the system sizes where quantum dynamics cannot be simulated using semiclassical or noise-added classical approximations. Remarkably, we find that a parametrically invariant meta-attractor emerges at a specific length scale and noise-added classical models deviate strongly from quantum dynamics below this scale. Our findings also generalize the previous surprising result that classically regular orbits can have the greatest quantum-classical differences in the semiclassical regime.  In particular, we show that the dynamical growth of quantum-classical differences is not determined by the degree of classical chaos. 
  

\end{abstract}

\maketitle
\setcounter{secnumdepth}{1}
\section{Introduction}
\label{sec:introduction}

Non-linearity is a quantum-mechanical resource useful in amplifying quantum-classical differences
\cite{jacobsEngineeringGiantNonlinearities2009a}. Quantum systems at the length and energy scales of current experimental
relevance are best treated as open to the environment and are thus termed NISQ (noisy intermediate scale quantum) 
systems \cite{preskillQuantumComputingNISQ2018}. These environmental effects lead to decoherence but can also be
exploited through measurement feedback \cite{eastmanTuningQuantumMeasurements2017,shiEffectsAmplificationFluctuation2019}. Experimentalists and theorists have both been interested in
understanding the emergence of quantum behavior in such non-linear quantum dissipative systems (NQDS)
\cite{bakemeierRouteChaosOptomechanics2015,liExperimentalSignaturesQuantumclassical2012}. Specifically, chaos, or sensitivity to initial conditions
and parameters, in an NQDS is fundamentally different from Hamiltonian chaos \cite{yusipovQuantumLyapunovExponents2019, ghoseTransitionClassicalChaos2004,kumarimeenuQuantumClassicalCorrespondenceEntanglement2019} and
can be quantified using the Lyapunov exponent.  In this context several groups including Pokharel et. al. \cite{pokharelChaosDynamicalComplexity2018},
Eastman et. al. \cite{eastmanTuningQuantumMeasurements2017}, Yusipov et. al. \cite{yusipovQuantumLyapunovExponents2019}, Ralph
et. al. \cite{ralphMultiparameterEstimationQuantum2017} have shown that Lyapunov exponents for NQDS can have a 
smooth \cite{ottEffectNoiseTimeDependent1984, dittrichQuantumEffectsSteady1987, klappaufObservationNoiseDissipation1998, klappaufErratumObservationNoise1999,
ammannQuantumDeltaKickedRotor1998, pattanayakParameterScalingDecoherent2003, gongQuantumChaosMeets2005, habibDecoherenceChaosCorrespondence1998} but non-monotonic
and rich quantum-to-classical transition. The details of the mechanism, including the role of the environment 
in this transition, have not been fully characterized.

In this context, we report on results from a study examining with care the effect of changing parameters on the
complexity of the dynamics. We consider semiclassical approximations
for the quantum Duffing oscillator that are derived from a quantum master equation under the assumption of
a Markovian environment. In the process we find that visual correspondence between different phase-space Poincare sections does not translate
to them having equal Lyapunov exponent and vice versa. Spatial and temporal complexity can be and are
independent of each other (see \cref{fig:poincare}). Therefore, we further analyze our semiclassical models using a Kullback-Liebler
distance between phase space attractors as well as the corresponding Lyapunov
exponents. Our results can be summarized as follows:
\begin{itemize}
    \item The primary effect of decreasing length scales is seen to be the increasing sensitivity to environmental
      fluctuations for systems of smaller size. However, this behavior is not explained by the effects of adding Gaussian
      noise to the classical system. Artificial noise-added classical systems reproduce some of the qualitative
      behavior of quantum noise temporally, but are measurably spatially different.
    \item At a certain length-scale the environment washes out difference between classically chaotic 
        and periodic trajectories. Consequently, a chaotic meta-attractor emerges which is invariant to changes 
        in environmental coupling. For systems that are smaller than this length-scale, 
        noise-added classical models diverge significantly from the semiclassical dynamics.
    \item The deviation from the classical limit is maximal for those systems where the global attractor is a classical periodic orbit (PO). 
        In other words, in keeping with previous results \cite{pokharelChaosDynamicalComplexity2018} these POs are more sensitive to environmental effects 
        of changing length scales than classically chaotic orbits and quantum-classical
      correspondence requires larger length-scales for POs than for chaos.
    \item In contrast to the standard intuition, even for chaotic systems the degree of classical chaos i.e. 
        the maximal classical Lyapunov exponent is {\em not} correlated
      to the length-scale where the classical and semiclassical dynamics deviate.
\end{itemize}

Thus, the careful consideration of Lyapunov exponents and the use of other measures reveal that the quantum to classical transition for
NQDS is filled with a wealth of non-intuitive phenomena arising from the interplay between chaos, noise and
quantum-effects at this scale. In what follows, we present our basic models and methods, before turning to results and a concluding discussion.
 
\section{Methods} 
\label{sec:methods}
The Duffing oscillator is a paradigmatic model to study
quantum to classical transition for chaos. The Newtonian model consists of a unit mass in a double-well potential with
dissipation \(\Gamma\), sinusoidal driving amplitude \(g\), frequency \(\omega\), and dimensionless
length scale \(\beta\),
\begin{equation}
\ddot{x} + 2 \Gamma \dot{x} +\beta^2x^3 -x = \frac{g}{\beta}cos(\Omega t). 
\label{eq:classical}
\end{equation}
The quantum Duffing oscillator is described using Quantum State Diffusion theory \cite{percivalQuantumStateDiffusion1998} 
where the stochastic evolution of a single pure quantum system \(|\psi\rangle\) under continuous measurement (including by the environment, for example) 
is considered. For this open system evolution, the Hamiltonian is
\begin{equation}
  \hat{H} = \frac{1}{2}\hat{P}^2 + \frac{\beta^2}{4}\hat{Q}^4 - \frac{1}{2}\hat{Q}^2 + \frac{\Gamma}{2}(\hat{Q}\hat{P}+\hat{P}\hat{Q})  -\frac{g}{\beta}\hat{Q}\cos(\Omega t),
\end{equation}
and dissipation due to coupling to the environment is represented by a single Lindblad operator
\(\hat{L} = \sqrt{\Gamma} (\hat{X} + i \hat{P})\) \cite{brunQuantumChaosOpen1996}. The corresponding Langevin-Ito equation for the evolution of the wavefunction is then  
\begin{equation} 
\begin{split}
\label{eq:SSE}
    | d\psi \rangle & = -\frac{i}{\hbar} \hat{H}  | \psi  \rangle dt + \sum_j(\hat{L_j}-\langle \hat{L_j} \rangle)  | \psi  \rangle d\xi_j + \sum_j(\langle \hat{L}^\dagger_j \rangle \hat{L}_j \\ & - \frac{1}{2} \hat{L}^\dagger_j\hat{L}_j - \frac{1}{2} \langle \hat{L}^\dagger_j \rangle \langle \hat{L}_j \rangle) | \psi  \rangle dt.
\end{split}
\end{equation}
In this system, both the dissipation and the quantum non-linearity scale with \(\Gamma\), via
\(\hat{L}\). The stochastic nature of the dynamics arises from independent normalized complex differential random variables
$d \xi_j$, where the mean over realizations $M$ satisfies
\(M(d\xi_j) = 0\ , M(d\xi_jd\xi^\prime_{j}) = 0\ , M(d\xi_j d\xi^{\prime*}_{j}) = \delta_{j\prime
  j}dt\).

This formulation of the Duffing oscillator has a single parameter
$\beta \equiv \sqrt{\frac{\hbar}{l^2 m \omega_0}}$ that determines the scale the system. The dynamics of
\cref{eq:classical}, defines the `classical limit', which is invariant under change in \(\beta\) except
for change of the length-scale. However, the dynamics of \cref{eq:SSE} do vary with \(\beta\); in particular \cref{eq:classical} 
can be derived for the expectation values of the position $\hat{X}, \hat{P}$ for the evolving \(| \psi \rangle\) as 
\(\beta \rightarrow 0\). Thus, this parameter allows us to study the transition from the quantum scale \(\beta \rightarrow 1\) to $\beta \to 0$, the largest length
scales where the quantum predictions become scale invariant and agree identically with the Newtonian predictions. In other words,
\(\beta\) represent the level of `quantumness' of the system. For our investigation, we focus on non-trivial
damping $0<\Gamma<0.35$ with $g = 0.3$, $\Omega = 1$ as in \cite{pokharelChaosDynamicalComplexity2018}, with an interval of
\(\delta \Gamma = 0.002\).

Previous work including that of Pokharel et al. \cite{pokharelChaosDynamicalComplexity2018} used a semiclassical stochastic equation
under the assumption that the wave function is sufficiently sharply localized by the action of the environment. In particular, this 
description reduces the dynamics to that of the first and second order moments of the
position and the momentum,
\(x = \langle \hat{X}\rangle, p = \langle\hat{P}\rangle, \mu = \sigma_{XX}, \kappa = \sigma_{PP}, R = \frac{1}{2} (\sigma_{XP} +
\sigma_{PX})\) where
\(\sigma_{AB} = \langle (A^{\dagger}- \langle A \rangle^*)(B - \langle B \rangle) \rangle\). It is possible to 
reduce this system further -- decreasing the number of variables in this stochastic ODE from five to four (see \cref{sec:appendix}) --
by observing a conserved quantity $\mu\kappa - R^2 = \frac{1}{4}$ and making a change of variables
\( \mu= \rho^2, R= \rho\Pi\). This yields the following semiclassical equations for the
Duffing oscillator:
\begin{equation}
 dx = pdt + 2 \sqrt{\Gamma} \bigg( \bigg( \rho^2 - \frac{1}{2} \bigg) d\xi_R - \rho \Pi d\xi_I \bigg),
\label{eq:SCdx}
\end{equation}
\begin{equation}
\begin{split}
  dp & = \bigg(-\beta^2x^3 + (1 -3 \beta^2 \rho^2)x - 2\Gamma p  + \frac{g}{\beta} \cos(\omega t) \bigg) dt 
   \\ & + 2 \sqrt{\Gamma}\bigg(\rho\Pi d\xi_R - \bigg ( \frac{1}{2} - \Pi^2 - \frac{1}{4\rho^2} \bigg) d\xi_I \bigg),
  \label{eq:SCdp}
\end{split}
\end{equation}
\begin{equation}
 \frac{d\rho}{dt} = \Pi + \Gamma \bigg( \rho - \rho^3 - \rho\Pi^2 + \frac{1}{4\rho} \bigg),
 \label{eq:SCdrho}
\end{equation}
\begin{equation}
  \frac{d\Pi}{dt} = \rho(-3\beta^2x^2+1) + \frac{1}{4\rho^3} - \Gamma\Pi \bigg( 1 + \Pi^2 + \rho^2 + \frac{3}{4\rho^2} \bigg).
  \label{eq:SCdpi}
\end{equation}
This system can be understood as the $x, p$ oscillator coupled to the $\rho$, $\Pi$ oscillator, with the latter
two variables representing the spread variables. The classical system is recovered from the semiclassical
approximation by increasing the system size i.e. by letting $\beta \rightarrow 0$. The semiclassical model
shows quantum features in three ways: 1) the influence of the spread variables \(\rho\) and \(\Pi\), 2) the environmental coupling
appearing via noise terms proportional to \(2 \sqrt{\Gamma}\) for \(dx\) and \(dp\), and 3) a tunneling-like effect
where the size of the barrier between the two wells of the oscillator is proportional to
\((1-3\beta^2\rho^2)\). In fact, this model can be described using a two-dimensional oscillator with the 
potential~\cite{shiEffectsAmplificationFluctuation2019} 
\begin{widetext}
\begin{equation}
\begin{split}
   U(x,\rho) = -(1-3\beta^2\rho^2)\frac{x^2}{2} + \beta^2(\frac{x^2}{2})^2 - (\frac{g}{\beta}\cos\omega t)x - \frac{\rho^2}{2} + \frac{1}{8\rho^2}
\end{split}
\end{equation}
\end{widetext}
that is also subject to noise and dissipation. In the remainder of the paper, we refer to this semiclassical model as ``SC.''

Recall that Eq.~(1) is \(\beta\) invariant other than a change in the length scales. Therefore, the natural
scale of analysis is \(x \beta, p \beta\). Under these `scaled' coordinates, the stochastic term in the
semiclassical Eqns. (4,5) scale with \(\beta\), while the tunneling term \(1 - 3 \beta^{2} \rho ^{2}\)
scales with \(\beta^{2}\). 
It is therefore likely that the initial effect of increasing \(\beta\) away from the classical limit arises
from the stochastic terms.  To isolate the effect of noise, we create a third model based on SC
by fixing $\beta = 10^{-5}$ -- a nearly classical length-scale where the semiclassical dynamics overlaps with
Newtonian dynamics -- and allowing for a scaling of the noise terms by a factor \(\beta_n\) so that
\(d \xi_{R,I} \rightarrow \beta_{n} d \xi_{R, I}\). We refer to this complex-noise added classical model as ``\(\text{C} + \mathcal{N}_{\mathbb{C}}\)''.
If our hypothesis about the role of the environment is correct, varying \(\beta_n\) for \(\text{C} + \mathcal{N}_{\mathbb{C}}\) should reproduce the effect
of varying \(\beta\) for SC. This immediately raises the question of whether a Gaussian noise-added classical model can reproduce the semiclassical
dynamics without having to go through the semiclassical dynamics. To this end we consider the Duffing oscillator with a 
noise term \(2 \sqrt{\Gamma} d \xi\) added to the classical Duffing equations, resulting in
\begin{equation} \label{noisyClassicalEqn1}
dx = pdt + 2 \sqrt{\Gamma} d\xi
\end{equation}
\begin{equation}
\label{noisyClassicalEqn2}
dp = -\beta^2x^3 + x - 2\Gamma p + \frac{g}{\beta} \cos(\omega t) dt + 2 \sqrt{\Gamma}d\xi.
\end{equation}
This real-noise added model is referred to as  \(\text{C} + \mathcal{N}_{\mathbb{R}}\).

Having established our various models to study the quantum-classical transition, we now turn to the quantitative metrics used
to characterize how the dynamics 
change as a function of system size. For example, the standard tool for characterizing complex dynamics via measuring sensitivity to initial conditions is the
largest Lyapunov exponent ($\lambda$) which distinguishes chaotic ($\lambda>0$) and periodic
($\lambda \leq 0$) dynamics \cite{boffettaPredictabilityWayCharacterize2002}. We calculate \(\lambda\) using the canonical
methods \cite{wolfDeterminingLyapunovExponents1985}. However, since the Lyapunov exponent of POs for the classical Duffing
oscillator at a given $\Gamma$ is \(\lambda_{\text{PO}} =-\Gamma\), we also use the previously introduced \cite{pokharelChaosDynamicalComplexity2018} dynamical complexity
\(K=\lambda+\Gamma\) to quantify how much more complex a given orbit is compared to the minimum case of a PO. $K$ is
strictly non-negative and compensates for the natural phase-space volume shrinkage caused by increasing
dissipation $\Gamma$ (see \cref{fig:lyap}). As a result, here $K$ allows for an effectively binary classification
of Duffing attractors (see \cref{fig:lyap}) across \(\beta\) and \(\Gamma\) with complex chaotic attractors having
\( 0.2 < K < 0.3 \) and simple periodic attractors having $K < 0.2$, independent of $\Gamma$. 

Despite their widespread and standard use, Lyapunov exponents can miss nuances
in the underlying dynamics.  In particular, the Lyapunov exponent quantifies the average temporal complexity
and not the spatial complexity.  The standard side-by-side comparison of Poincare sections, in this case for
different semiclassical approximations (as shown in \cref{fig:poincare}), can be somewhat informative in understanding spatial complexity differences. However, different stochastic terms
(\cref{eq:SCdx,eq:SCdp,eq:SCdrho,eq:SCdpi}) ``blur'' the Poincare sections in different ways that are not
visually distinguishable. In other words, visual comparisons do
not allow us to quantify the differences between dissimilar Poincare sections or to rank-order dissimilarity. To this end, we introduce a  Kullback-Leibler inspired spatial similarity metric
\(d\) that allows us to compare different phase-space Poincare sections. In particular, for two histogram distributions \(f_1(x)\)
and \(f_2(x)\), the distance \(l\)  \cite{pattanayakParameterScalingDecoherent2003}  is defined as 
\begin{equation}
 \label{eq:sKL}
 l(f_{1}, f_{2}) = -  \ln \left[ \frac{\int \Big(f_{1}(x) \cdot f_{2}(x)\Big)^{2} \mathrm{d}x}{ \Big ( \int f_{1}(x) \cdot f_{1}(x) \mathrm{d}x \Big) \Big( \int f_{2}(x) \cdot f_{2}(x) \mathrm{d}x \Big)} \right]. 
\end{equation}
Given a dynamical model \(M\) (semi-classical or noise-added classical), we compute the \(l\) over each coordinate of the phase space and take the
Eulerian sum
\begin{equation}
d_{M_{1}, M_{2}}(\Gamma_{1}, \beta_{1}; \Gamma_{2}, \beta_{2}) =  \sqrt{\sum\limits_{j \in \{ x,p \}} l \left(M_{1}^{j}(\Gamma_{1}, \beta_{1}), M_{2}^{j}(\Gamma_{2}, \beta_{2})\right)^{2}}
\end{equation}
where \(M^{j}\) is the coarse-grained Poincare-section histogram for the \(j\)-th phase-space coordinate. Note that we do not
consider the coordinates \(\rho, \Pi\) as they are only defined for SC and \(\text{C} + \mathcal{N}_{\mathbb{C}}\).
Two identical Poincare sections yield \(d=0\) and two orthogonal sections yield \(d \rightarrow \infty\). This
distance is only well-defined for chaotic trajectories; for periodic trajectories, the distributions are too
localized and even slight differences can lead to large distances (see \cref{sec:d_divergence} for more details).

\section{Results and Discussion}
\label{sec:results}
\begin{figure}
    \centering
    \includegraphics[width=\linewidth]{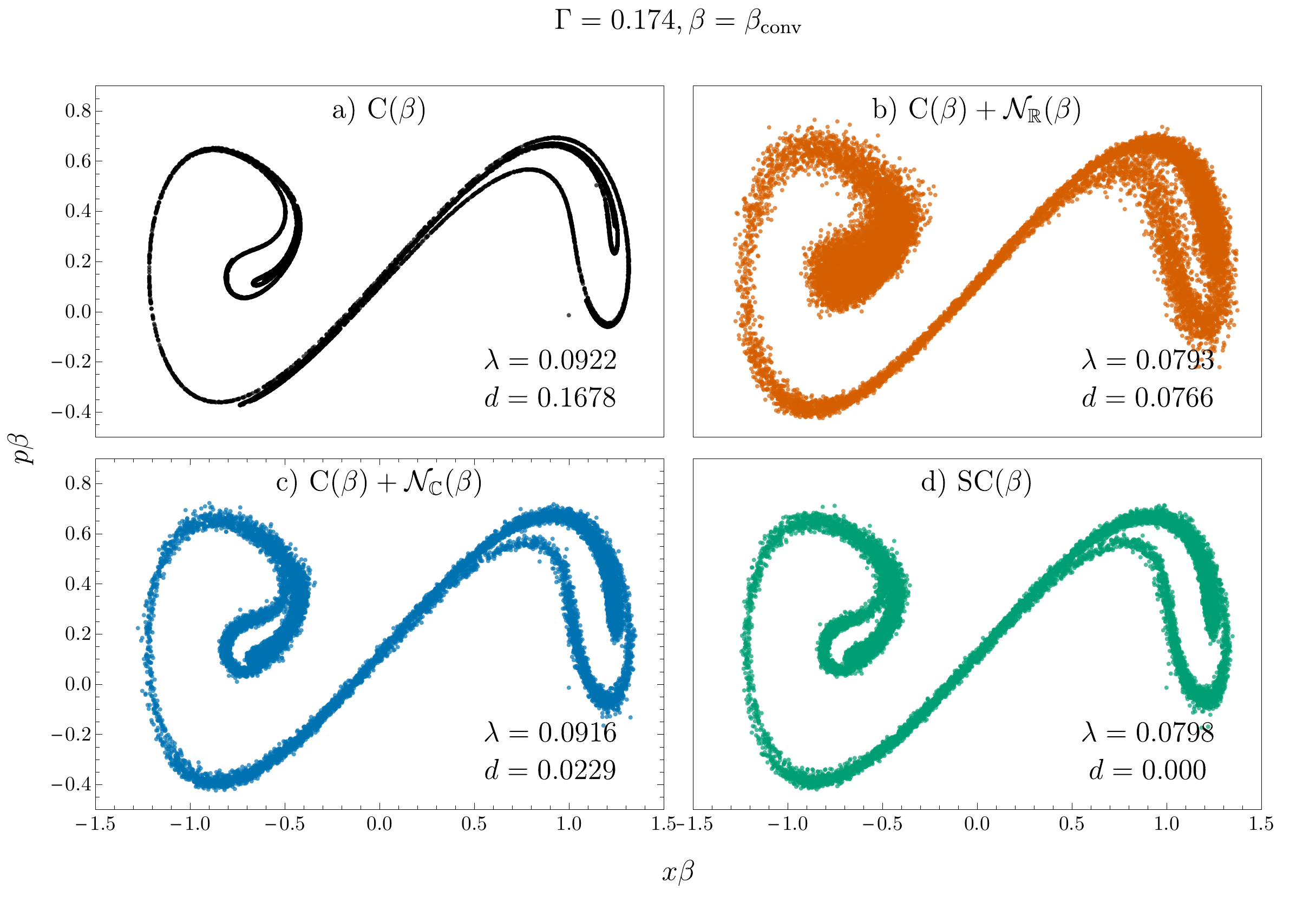}
    \caption{Different approximations are compared to the semiclassical model. Chaotic attractors for the
      Duffing oscillator at a representative environmental coupling ($\Gamma = 0.174$) and system size
      ($\beta = \beta_{\text{conv}}$) are shown. The four figures are obtained under different semiclassical
      approximations described in more detail in the text. Note that while \hyperref[fig:poincare]{(c)} and \hyperref[fig:poincare]{(d)} are hard to distinguish
      visually, they are well-separated using the Lyapunov exponent or the SKL measure. At the same time,
      attractors in \hyperref[fig:poincare]{(a)} and \hyperref[fig:poincare]{(c)}, and \hyperref[fig:poincare]{(b)} and \hyperref[fig:poincare]{(d)} have the same Lyapunov exponent respectively.  All four approximations have a unique distance $d$ from SC, with cruder approximations appropriately having
      a larger $d$. }
    \label{fig:poincare}
\end{figure}

 \begin{figure*}[t!]
         \includegraphics[width=\linewidth]{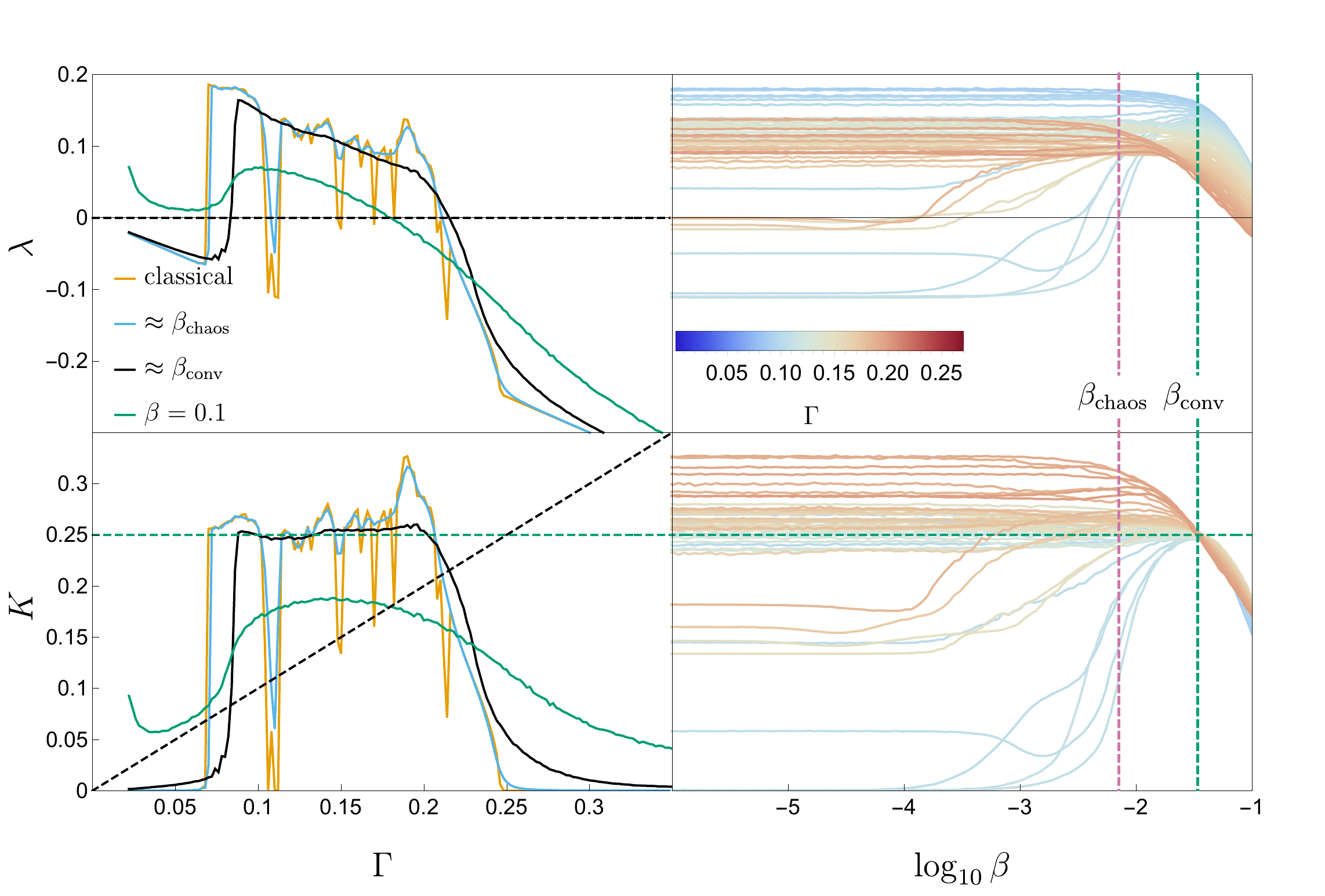}
         \caption{The Lyapunov exponent ($\lambda$) and the dynamical complexity ($K$) as a function
           of dissipation $\Gamma$ and system size $\beta$ are shown. The right two plots focus on
           $0.088 < \Gamma \leq 0.2$. Both plots show that all classically periodic orbits become
           chaotic at $\beta = \beta_{\text{chaos}}$. Classically chaotic dynamics, in comparison,
           are robust as the length scales decreases. Note also that there is a convergence of the
           dynamical complexity at $\beta_{\text{conv}}$. This convergence is also confirmed using
           $d$ in \cref{fig:clusterKL}.}
        \label{fig:lyap}
      \end{figure*}

      \begin{figure}
       \centering
         \includegraphics[width=0.8\linewidth]{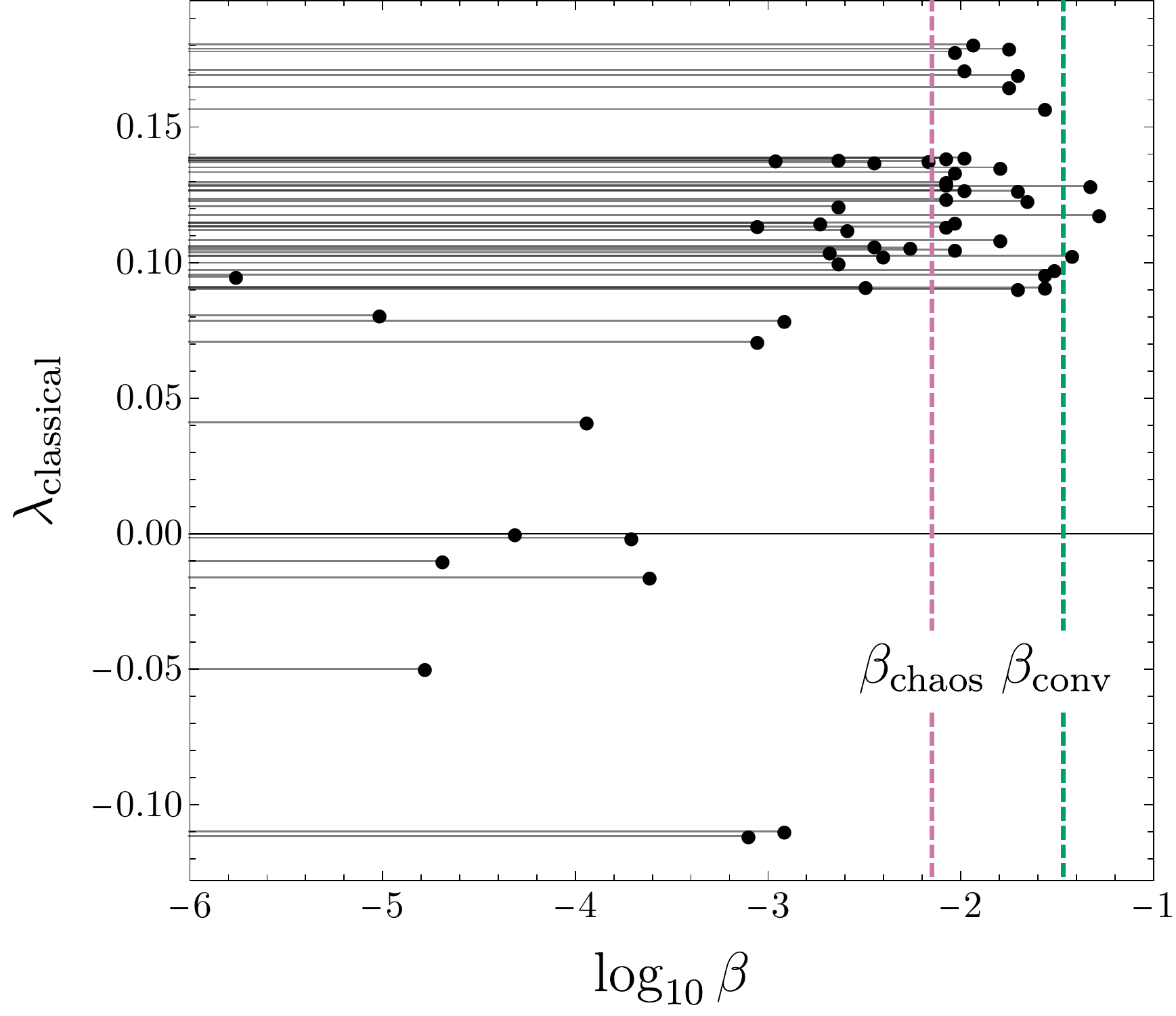}
  \caption{\( \beta_{\text{break}} \) is the length scale at which \( \lambda(\beta) \) diverges from the
    classical Lyapunov. \(\beta_{\text{break}}\) as a function of \(\lambda_{\text{classical}}\) is
    shown.}
    \label{fig:breaktime_sc}
 \end{figure}

\begin{figure}[ht]
         \includegraphics[width=0.7\linewidth]{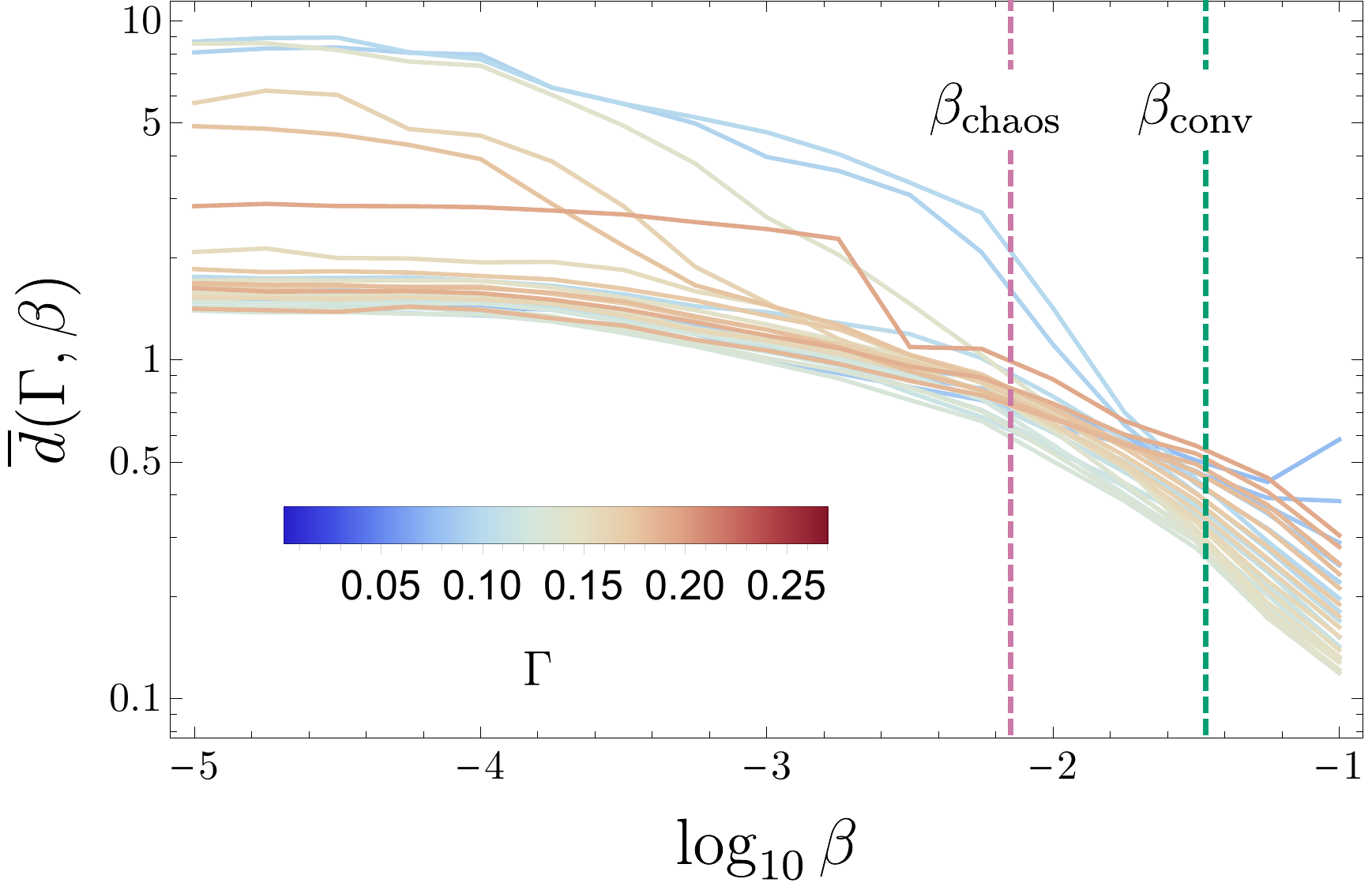}
         \caption{The mean distance between a given oscillators at some \(\Gamma\) and all its counterparts in
           the range $[0.088, 0.2]$ is shown. Corroborating the observation in \cref{fig:lyap}, this
           distance between all attractors is decreasing for all oscillators, with all attractors becoming
           indistinguishable from the meta-attractor as the length-scale decreases. }
        \label{fig:clusterKL}
\end{figure}

\begin{figure}[ht]
  \includegraphics[width=\linewidth]{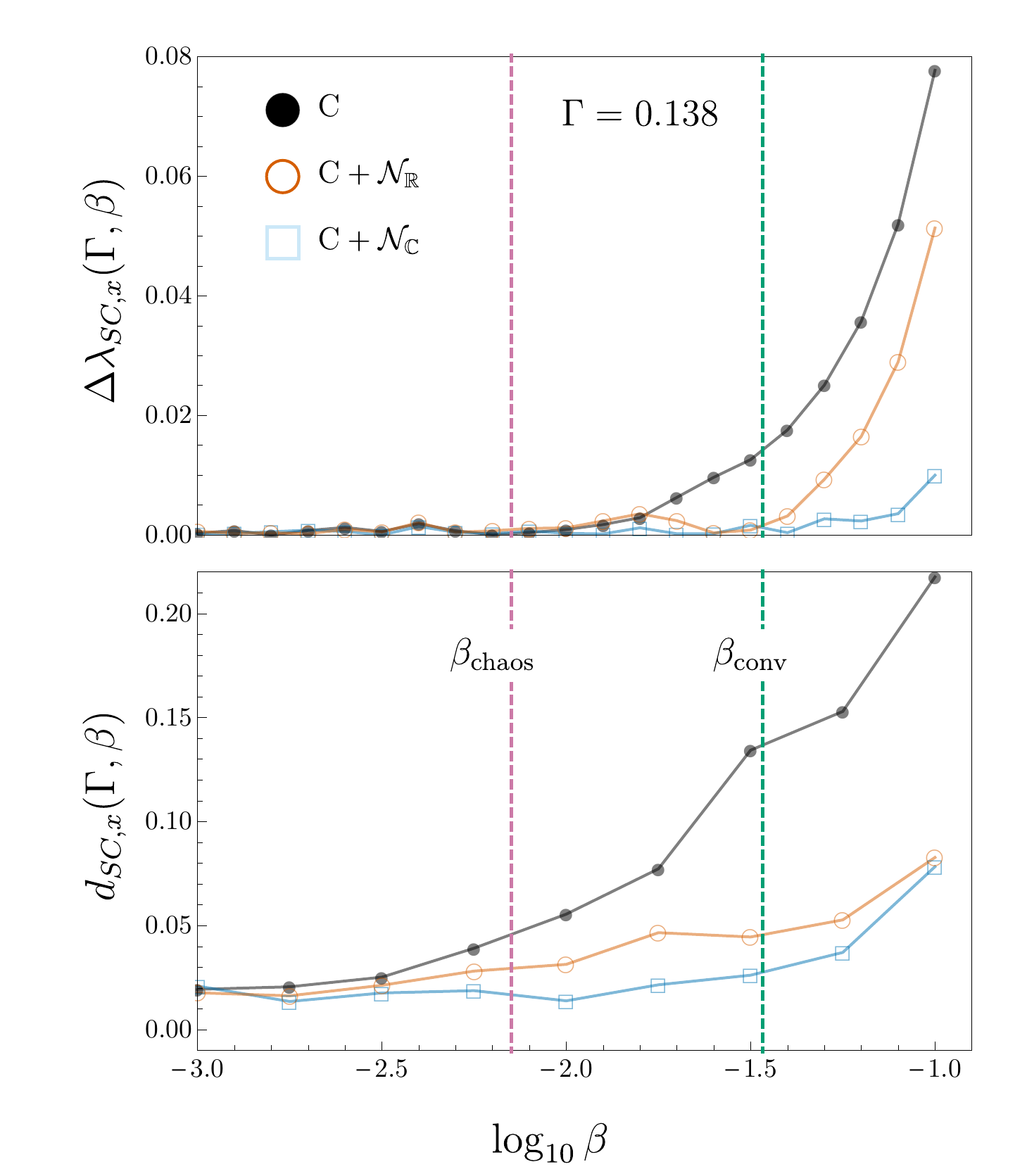}
  \caption{The change in difference between semiclassical approximations as a function of system
    size as captured by $\Delta \lambda$ and $d$. The oscillator chosen is the central oscillator,
    which has $\Gamma = 0.138 $. For this $\Gamma$ both metrics show a monotonic increase in the
    difference as the system size gets smaller and quantum effects become more prominent. However,
    the difference in dynamics at a larger system sizes is visible in $d$, than in $\lambda$. In
    other words, there exist length scales (for instance $\beta = \beta_{\text{chaos}}$) for which
    approximate models do not reproduce the spatial attractor, even though the Lyapunov exponent is
    reproduced.}
    \label{fig:klyapint}
\end{figure}

\begin{figure*}[ht]
  \includegraphics[width=\linewidth]{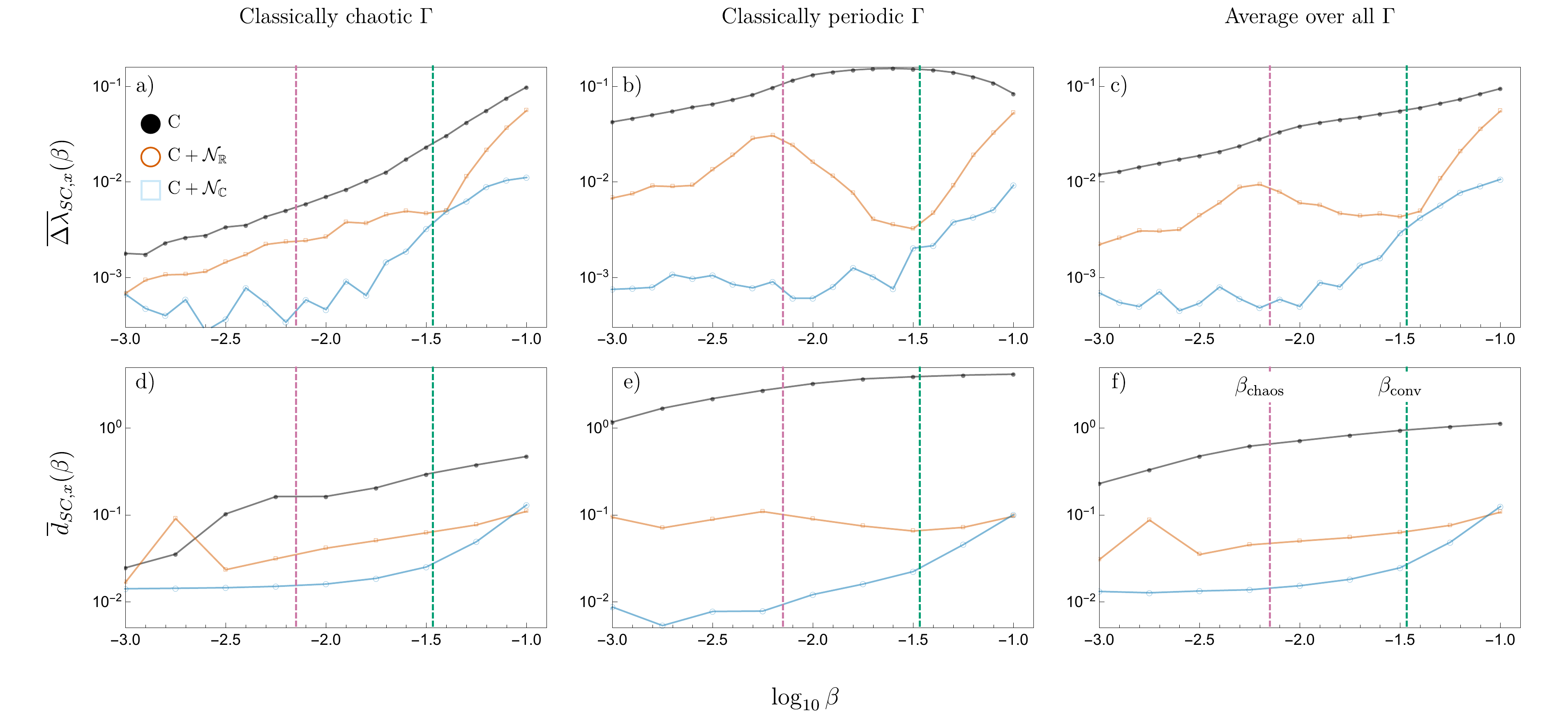}
 \caption{  $\Delta \lambda$ and $d$ shown as a function of length scale $\beta$, while averaged over various $\Gamma$ ranges. Even though different approximations to the semiclassical are visually (in terms of Poincare sections) and qualitatively similar to the full semiclasssical model, the difference between the models can quantified.  Just like \cref{fig:klyapint}, both difference measures are monotonic in \hyperref[fig:lyapKLgrid]{(a)} and \hyperref[fig:lyapKLgrid]{(d)} for classically chaotic attractors, other than an anomalous peak at $\log_{10} \beta = -2.75$ (see Appendix for more details). On the other hand, classically periodic orbits are the hardest to recover using approximate semiclassics: the non-monotonic behavior of $\Delta \lambda$ over the entire range of $\Gamma$ is due to the classically periodic orbits in \hyperref[fig:lyapKLgrid]{(b)} and \hyperref[fig:lyapKLgrid]{(e)}.$\Delta \lambda$ and $d$ between the semiclassical and noise-added classical models is the highest at $\beta = \beta_{\text{chaos}}$, after which all oscillators become chaotic. At $\beta = \beta_{\text{conv}}$, noise washes out the differences between different oscillators leading to the convergence shown in Fig2and3; and all the approximate semiclassical models lose relevance for $\beta < \beta_{\text{conv}}$.  }

 \label{fig:lyapKLgrid}
\end{figure*}

At the classical limit, the Duffing oscillator has a standard bifurcation transition between regularity and complexity 
as a function of most dynamical parameters. Changing the strength $\Gamma$ of the coupling to the
environment (which classically affects only the dissipation due to the environment and not fluctuations) also 
induces this behavior, including windows of periodic behavior and a regime of co-existing attractors. This parametric 
sensitivity which leads  to abrupt changes in dynamical behavior has potential consequences in metrology as well applications in quantum 
control~\cite{eastmanTuningQuantumMeasurements2017}.
Before we proceed to map the details of how the variety of classical dynamics and the classical sensitivity to parameters behave
under change of length scale, we first establish the value of independently analyzing temporal complexity (using \(\lambda\)) and
spatial similarity (using \(d\)). To do this, consider a classically chaotic attractor at \(\Gamma = 0.174\) under
different approximations \(\text{C}, \text{C} + \mathcal{N}_{\mathbb{R}} \text{, and } \text{C}  + \mathcal{N}_{\mathbb{C}}\). In
    \cref{fig:poincare} we compute \(\lambda\) and the distance \(d_{SC,x}\) for different semiclassical approximations. We find that:
    \begin{itemize}
    \item phase-space attractors that look different (like Fig. \hyperref[fig:poincare]{1(b)} and \hyperref[fig:poincare]{1(d)}) are quantifiably more similar to each other in
      \(d\) than ones that at first appearance look similar (Fig. \hyperref[fig:poincare]{1(a)} and \hyperref[fig:poincare]{1(d)}).
    \item statistical differences between the semiclassical (SC) and classical-noise added model
      (\(\text{C}
      + \mathcal{N}_{\mathbb{R}}\)), which is missed by \(\lambda\), do produce quantitatively and
      visually different spatial complex Poincare sections (Fig. \hyperref[fig:poincare]{1(b)} and \hyperref[fig:poincare]{1(d)}),
    \item even though the phase-space attractors for \(\text{C} + \mathcal{N}_{\mathbb{C}}\) and SC are visually
      nearly identical, there are differences that can be captured by using \(d\) (Fig. \hyperref[fig:poincare]{1(c)} and \hyperref[fig:poincare]{1(d)}),
\end{itemize}

With these caveats in mind, consider the temporal complexities \(\lambda(\Gamma)\) and
\(K(\Gamma)\).  The \(\Gamma\) landscape can be divided classically into three regimes: low ($\Gamma < 0.068$),
intermediate ($0.068 \leq \Gamma < 0.2$) and high (\(\Gamma > 0.2\)). In the low-damping regime, the orbits are
regular orbits that traverse both wells of the Duffing oscillator; while in the high-damping regime, owing to
the overwhelming magnitude of \(\Gamma\) we observe single-well regular orbits.

Motivated by the inherent linearity of Hamiltonian quantum dynamics, it has been suggested that quantum
mechanics makes systems more regular. Further, as a result, the quantum-classical difference are argued to scale with 
the Lyapunov exponent of the classical dynamics, with the classically chaotic systems having 
maximal deviation from quantum behavior. However, in NQDS in general and \cref{fig:lyap} in
particular, we see all four possible transitions as a function of \(\beta\): regular-to-regular,
regular-to-chaos, chaos-to-regular and chaos-to-chaos. For example,
the lower portion of the intermediate regime ($0.068 \leq \Gamma < 0.088$) shows a transition from classical
chaos to semiclassical regularity. Here quantum effects delay the impact of increasing dissipation. Therefore,
we see a `quantum regularization' of chaos and the increase in \(\lambda\) as a function of \(\Gamma\) gets
shifted. Some of the interesting effects due to the existence of a co-existing attractor in this regime are 
discussed further in \cref{sec:low_and_high}. The bulk of our analysis here, however,
focuses on the intermediate regime $0.088 \leq \Gamma < 0.2$. Here we see chaos-to-chaos and
regular-to-chaos transitions as a function of \(\beta\).

To further clarify the effect of decreasing effective system action consider the behavior of temporal
complexity as function of \(\beta\) in \cref{fig:lyap,fig:breaktime_sc}. As
Pokharel et. al. \cite{pokharelChaosDynamicalComplexity2018} previously noted, introducing more quantum
effects (increasing $\beta$) decreases the parametric sensitivity of temporal complexities to $\Gamma$ when compared
to the classical system. This `chaotificiation' is evident in the 
smoothing of the jaggedness of $K(\Gamma)$ (where the jaggedness corresponds to dips of regularity arising from higher order POs) 
as we scan \(\beta > \beta_{\text{classical}}\). This can be
understood as arising primarily because the addition of `quantum' terms to the classical equations quickly
destroys the higher period POs, which rely on dynamical synchronization across many periods. Consequently, all periodic attractors for
\(\Gamma \in (0.088, 0.2] \) become chaotic for scales below \(\beta_{\text{chaos}}=
0.0068\). Moreover, the temporal complexity curves \(\lambda(\beta), K(\beta)\) are
highly non-monotonic and idiosyncratic as a function of \(\beta\) (see \cref{fig:lyap}). This is clearly a
distinctive feature of NQDS. 

A re-plotting of temporal complexity with a focus at the length scales where the
classical dynamics no longer agrees with the semiclassical dynamics (i.e. \(\beta_{\text{break}}\)) is shown in
\cref{fig:breaktime_sc}. Here not only is \(\beta_{\text{break}}\) lower for classical POs but also the
break-scale is not intuitively related to the degree of chaos
in the classical system \(\lambda_{\text{classical}}\). This is in contrast to the standard arguments about
quantum `break time' that have persisted in the literature for many decades~\cite{bermanConditionStochasticityQuantum1978a} but 
were constructed without considering the fact that genuine quantum-classical correspondence is only obtained by considering the effect 
of decoherence via the environment.

Further, while \(\lambda\) and \(K\) both capture \(\beta\)-invariance, only \(K\) captures that the complexity
of the dynamics becomes invariant of environmental coupling \(\Gamma\) at a certain length scale. To be
more specific,  at a particular degree of quantumness \(\beta = \beta_{\text{conv}} = 0.0341\),  \(K (\Gamma)\)
flattens. This remarkable convergence in dynamical complexity is verified by looking at the spatial similarity between
the phase-space attractors. In particular, in \cref{fig:clusterKL}, we compute the average
distance between a trajectory at \(\Gamma\) to all its counterparts in the range \(\Gamma \in (0.088, 0.2] \)
under the same model \(M\) and system size \(\beta\)
\begin{equation}
\overline{d}(\Gamma, \beta) = \text{mean}_{\Gamma'} d_{M, M} (\Gamma, \beta; \Gamma', \beta). 
\end{equation}
This average spatial similarity between attractors decreases as a function of system size. In 
other words, near \(\beta_{\text{conv}}\) both spatial and temporal complexity are
invariant under changes in the environmental coupling.

To summarize, the competition between environmental effects and chaos is richest in between
\(\beta_{\text{classical}} \leq \beta \leq \beta_{\text{conv}}\), the upper and lower boundaries of which have
\(\beta\) and \(\Gamma\) invariant attractors respectively.  As we detail below,
\(\text{C} + \mathcal{N}_{\mathbb{C}}\) closely mirrors semiclassical dynamics for systems larger than
\(\beta_{\text{conv}}\), which means that this range of scales is where noise from environmental coupling
plays a significant role. At the same time, other than qualitative differences in how classically chaotic and
regular orbits behave as a function of system size, there is no clear connection between the degree of 
classical chaos and quantum-classical deviation.

Now we turn from individual behavior to global characterization of various models. To do this, we compare the dynamics
under different models by averaging distances over different parameters. In particular, we define a measure of
the distance between two different models \(M_{1}\) and \(M_{2}\) at some specified system size \(\beta\) and
averaged over a range of \(\Gamma\) as:
\begin{equation}
\overline{d}_{M_1, M_2} (\beta) = \text{mean}_{\Gamma} \ d_{M_1, M_2}(\Gamma, \beta; \Gamma, \beta).
\end{equation}
An equivalent similarity metric can also be defined for comparing temporal complexity
\begin{equation}
\overline{\Delta \lambda}_{M_1, M_2}(\beta) = \text{mean}_{\Gamma} \left| \lambda_{M_1}(\Gamma, \beta) - \lambda_{M_2} (\Gamma, \beta) \right|.
\end{equation}

We start with a typical case of a classically chaotic attractor at \(\Gamma = 0.138\) in
\cref{fig:klyapint}. Here we see that: a) all of the approximations to the accurate semiclassics get monotonically
worse as the system size decreases; b) as we increase the number of semiclassical
terms, the difference between the noise-added models and the semiclassical model decreases i.e. that
\(\text{C} + \mathcal{N}_{\mathbb{C}}\) does better than  \(\text{C} + \mathcal{N}_{\mathbb{R}}\); 
c) comparing the top and the bottom figures,  we notice that even when
the Lyapunov exponents under different models are identical, for instance at \(\beta_{\text{chaos}}\), the
underlying attractors might not be spatially similar. As we discuss below,
b) and c) remain true in general for both classically chaotic and regular oscillators, but a) need not.

This simplicity does not persist in the aggregate behavior shown in \cref{fig:lyapKLgrid}. Here the figures in
third column are averaged over all \(\Gamma\) in the range \((0.088, 0.2]\) and are the weighted sum of the
first two columns, which are over the classically chaotic and periodic trajectories respectively. 
The aggregate behavior looks far more structured and to understand this structure, we first note that the rapid fluctuation in
\(\overline{\Delta \lambda}(\beta)\) for  \(\text{C} + \mathcal{N}_{\mathbb{C}}\) are of the order \(10^{-3}\) which are inherent numerical
precision errors arising from how the Lyapunov exponents are computed and further visually amplified by the
log-scale. Apart from the overall increasing deviation of noise-added models from the semiclassical model SC as a function of \(\beta\), there are
several non-monotonic features which we now discuss. 

The non-monotonicity of \(\overline{\Delta \lambda }(\beta)\) for the overall
average behavior in Fig. \hyperref[fig:lyapKLgrid]{6(c)} is due to the classical POs, as clearly seen in Fig.~\hyperref[fig:lyapKLgrid]{6(b)}. 
The peak of \(\Delta \lambda\) near \(\beta_{\text{chaos}}\) in Fig. \hyperref[fig:lyapKLgrid]{6(b)} highlights the sensitivity of
classical POs to Gaussian noise as this sensitivity drops when all trajectories are chaotic for
\(\beta > \beta_{\text{chaos}}\). Put differently, the semiclassical behavior of classical POs is harder to
estimate using classical-noise added models and consequently requires the more sophisticated dynamical description given by \(\text{C} + \mathcal{N}_{\mathbb{C}}\) and SC. 

The non-monotonicity in \(\overline{d}(\beta)\) is largely due to an anomalous peak at \(\log_{10} \beta = - 2.75\),
most visible in Fig. \hyperref[fig:lyapKLgrid]{6(d)} and  \hyperref[fig:lyapKLgrid]{6(f)}. At this \(\beta\)-value, we see a single-well chaotic
attractor in the noise-added models, which is spatially different from the true semiclassical double-well
attractor (see \cref{sec:d_divergence} for more details). This peak in \(\overline{d}(\beta)\) is not picked up by the Lyapunov exponents. 
Overall, we find that the spatial metric measuring the difference between phase-space attractors, \(d\), does better at distinguishing 
between different models than \(\lambda\). This emphasizes the need to use both spatial and temporal complexity metrics separately to study NQDS. 
Even when averaged, we see that spatial and temporal similarity of the models to the
semiclassical dynamics gets progressively better as we use more sophisticated approximations. Moreover, for \(\beta < \beta_{\text{conv}}\)
the \(\text{C}+\mathcal{N}_{\mathbb{C}}\) model approximates the semiclassical model well. This close similarity between SC and
\(\text{C}+\mathcal{N}_{\mathbb{C}}\) suggests that increased sensitivity to the environment is the dominant mechanism by which quantum
effects manifest in this NQDS.

\section{Conclusion}
\label{sec:conclusion}    

Nonlinear quantum dissipative systems evolve under an interplay between quantum effects, noise, and non-linearity. 
Our exploration of the quantum-classical transition in an NQDS, in particular a noisy nonlinear driven oscillator, 
shows that the semiclassical regime of NQDS remains an intriguing regime for further exploration. We demonstrate 
that while Lyapunov exponents and visual examinations of the phase space Poincare sections provide a reasonable first 
gauge of change in dynamical complexity or attractor similarity across
systems, more careful metrics provide a better and more nuanced insights about this transition. We have also investigated 
whether the semiclassical dynamics here can be explained by simply adding classical noise. We find that while classical 
noise produces certain qualitative features, there are spatial and temporal quantitative features that require taking 
quantum effects into account. Surprisingly, the length scale for the
breakdown of classical approximations to quantum behavior is not determined by the degree of
classical chaos in the system, and the growth of quantum-classical difference is in general quite
idiosyncratic. Such semiclassical techniques have been studied and used widely over the decades, particularly in 
chemical physics ~\cite{prezhdoClassicalMappingSecondorder2002,tomsovicSemiclassicalDynamicsChaotic1991,banerjeeSolutionQuantumLangevin2004}. 
Therefore, understanding what precisely does determine the break length scale 
is of great interest and remains an interesting avenue for further work. Further, we find that smaller systems that are more `quantum' have 
few complicated periodic orbits due to increased sensitivity to environmental fluctuations. They are thus less sensitive to parameter
variation than their classical counterparts. Remarkably, this leads to a chaotic meta-attractor
fairly deep in the quantum regime that exhibits dynamical complexity relatively independent of the length-scale
or environmental coupling. Arguably, even our most sophisticated semiclassical model, which implicitly uses an environment consisting of a 
Markovian bath at zero temperature, remains relatively simple. In particular, considering a finite bath at a finite temperature with 
manifestly non-Markovian features would allow for richer investigation of NQDS. Analyzing how features of the current analysis are altered
by different, more sophisticated noise spectra, is another obvious next challenge. 

\bigbreak

\section{Acknowledgments}
\label{sec:acknowledgements}
We would like to thank Bruce Duffy for computational support, and BP would like to thank Namit Anand for insightful discussions. AP is grateful for HHMI funding through Carleton, and internal funding from Carleton for student research. 
      
\bibliography{main, chaos_references}

\begin{thebibliography}{29}%
\makeatletter
\providecommand \@ifxundefined [1]{%
 \@ifx{#1\undefined}
}%
\providecommand \@ifnum [1]{%
 \ifnum #1\expandafter \@firstoftwo
 \else \expandafter \@secondoftwo
 \fi
}%
\providecommand \@ifx [1]{%
 \ifx #1\expandafter \@firstoftwo
 \else \expandafter \@secondoftwo
 \fi
}%
\providecommand \natexlab [1]{#1}%
\providecommand \enquote  [1]{``#1''}%
\providecommand \bibnamefont  [1]{#1}%
\providecommand \bibfnamefont [1]{#1}%
\providecommand \citenamefont [1]{#1}%
\providecommand \href@noop [0]{\@secondoftwo}%
\providecommand \href [0]{\begingroup \@sanitize@url \@href}%
\providecommand \@href[1]{\@@startlink{#1}\@@href}%
\providecommand \@@href[1]{\endgroup#1\@@endlink}%
\providecommand \@sanitize@url [0]{\catcode `\\12\catcode `\$12\catcode
  `\&12\catcode `\#12\catcode `\^12\catcode `\_12\catcode `\%12\relax}%
\providecommand \@@startlink[1]{}%
\providecommand \@@endlink[0]{}%
\providecommand \url  [0]{\begingroup\@sanitize@url \@url }%
\providecommand \@url [1]{\endgroup\@href {#1}{\urlprefix }}%
\providecommand \urlprefix  [0]{URL }%
\providecommand \Eprint [0]{\href }%
\providecommand \doibase [0]{http://dx.doi.org/}%
\providecommand \selectlanguage [0]{\@gobble}%
\providecommand \bibinfo  [0]{\@secondoftwo}%
\providecommand \bibfield  [0]{\@secondoftwo}%
\providecommand \translation [1]{[#1]}%
\providecommand \BibitemOpen [0]{}%
\providecommand \bibitemStop [0]{}%
\providecommand \bibitemNoStop [0]{.\EOS\space}%
\providecommand \EOS [0]{\spacefactor3000\relax}%
\providecommand \BibitemShut  [1]{\csname bibitem#1\endcsname}%
\let\auto@bib@innerbib\@empty
\bibitem [{\citenamefont {Jacobs}\ and\ \citenamefont
  {Landahl}(2009)}]{jacobsEngineeringGiantNonlinearities2009a}%
  \BibitemOpen
  \bibfield  {author} {\bibinfo {author} {\bibfnamefont {K.}~\bibnamefont
  {Jacobs}}\ and\ \bibinfo {author} {\bibfnamefont {A.~J.}\ \bibnamefont
  {Landahl}},\ }\href {\doibase 10.1103/PhysRevLett.103.067201} {\bibfield
  {journal} {\bibinfo  {journal} {Phys. Rev. Lett.}\ }\textbf {\bibinfo
  {volume} {103}},\ \bibinfo {pages} {067201} (\bibinfo {year}
  {2009})}\BibitemShut {NoStop}%
\bibitem [{\citenamefont {Preskill}(2018)}]{preskillQuantumComputingNISQ2018}%
  \BibitemOpen
  \bibfield  {author} {\bibinfo {author} {\bibfnamefont {J.}~\bibnamefont
  {Preskill}},\ }\href {\doibase 10.22331/q-2018-08-06-79} {\bibfield
  {journal} {\bibinfo  {journal} {Quantum}\ }\textbf {\bibinfo {volume} {2}},\
  \bibinfo {pages} {79} (\bibinfo {year} {2018})}\BibitemShut {NoStop}%
\bibitem [{\citenamefont {Eastman}\ \emph {et~al.}(2017)\citenamefont
  {Eastman}, \citenamefont {Hope},\ and\ \citenamefont
  {Carvalho}}]{eastmanTuningQuantumMeasurements2017}%
  \BibitemOpen
  \bibfield  {author} {\bibinfo {author} {\bibfnamefont {J.~K.}\ \bibnamefont
  {Eastman}}, \bibinfo {author} {\bibfnamefont {J.~J.}\ \bibnamefont {Hope}}, \
  and\ \bibinfo {author} {\bibfnamefont {A.~R.~R.}\ \bibnamefont {Carvalho}},\
  }\href {\doibase 10.1038/srep44684} {\bibfield  {journal} {\bibinfo
  {journal} {Sci. Rep.}\ }\textbf {\bibinfo {volume} {7}},\ \bibinfo {pages}
  {44684} (\bibinfo {year} {2017})}\BibitemShut {NoStop}%
\bibitem [{\citenamefont {Shi}\ \emph {et~al.}(2019)\citenamefont {Shi},
  \citenamefont {Greenfield}, \citenamefont {Eastman}, \citenamefont
  {Carvalho},\ and\ \citenamefont
  {Pattanayak}}]{shiEffectsAmplificationFluctuation2019}%
  \BibitemOpen
  \bibfield  {author} {\bibinfo {author} {\bibfnamefont {Y.}~\bibnamefont
  {Shi}}, \bibinfo {author} {\bibfnamefont {S.}~\bibnamefont {Greenfield}},
  \bibinfo {author} {\bibfnamefont {J.}~\bibnamefont {Eastman}}, \bibinfo
  {author} {\bibfnamefont {A.}~\bibnamefont {Carvalho}}, \ and\ \bibinfo
  {author} {\bibfnamefont {A.}~\bibnamefont {Pattanayak}},\ }in\ \href
  {\doibase 10.1007/978-3-030-10892-2_9} {\emph {\bibinfo {booktitle}
  {Proceedings of the 5th International Conference on Applications in Nonlinear
  Dynamics}}}\ (\bibinfo {organization} {Springer},\ \bibinfo {year} {2019})\
  pp.\ \bibinfo {pages} {72--83}\BibitemShut {NoStop}%
\bibitem [{\citenamefont {Bakemeier}\ \emph {et~al.}(2015)\citenamefont
  {Bakemeier}, \citenamefont {Alvermann},\ and\ \citenamefont
  {Fehske}}]{bakemeierRouteChaosOptomechanics2015}%
  \BibitemOpen
  \bibfield  {author} {\bibinfo {author} {\bibfnamefont {L.}~\bibnamefont
  {Bakemeier}}, \bibinfo {author} {\bibfnamefont {A.}~\bibnamefont
  {Alvermann}}, \ and\ \bibinfo {author} {\bibfnamefont {H.}~\bibnamefont
  {Fehske}},\ }\href {\doibase 10/f63q5r} {\bibfield  {journal} {\bibinfo
  {journal} {Phys. Rev. Lett.}\ }\textbf {\bibinfo {volume} {114}},\ \bibinfo
  {pages} {013601} (\bibinfo {year} {2015})}\BibitemShut {NoStop}%
\bibitem [{\citenamefont {Li}\ \emph {et~al.}(2012)\citenamefont {Li},
  \citenamefont {Kapulkin}, \citenamefont {Anderson}, \citenamefont {Tan},\
  and\ \citenamefont
  {Pattanayak}}]{liExperimentalSignaturesQuantumclassical2012}%
  \BibitemOpen
  \bibfield  {author} {\bibinfo {author} {\bibfnamefont {Q.}~\bibnamefont
  {Li}}, \bibinfo {author} {\bibfnamefont {A.}~\bibnamefont {Kapulkin}},
  \bibinfo {author} {\bibfnamefont {D.}~\bibnamefont {Anderson}}, \bibinfo
  {author} {\bibfnamefont {S.}~\bibnamefont {Tan}}, \ and\ \bibinfo {author}
  {\bibfnamefont {A.}~\bibnamefont {Pattanayak}},\ }\href {\doibase
  10.1088/0031-8949/2012/T151/014055} {\bibfield  {journal} {\bibinfo
  {journal} {Phys. Scripta}\ }\textbf {\bibinfo {volume} {T151}},\  (\bibinfo
  {year} {2012})}\BibitemShut {NoStop}%
\bibitem [{\citenamefont {Yusipov}\ \emph {et~al.}(2019)\citenamefont
  {Yusipov}, \citenamefont {Vershinina}, \citenamefont {Denisov}, \citenamefont
  {Kuznetsov},\ and\ \citenamefont
  {Ivanchenko}}]{yusipovQuantumLyapunovExponents2019}%
  \BibitemOpen
  \bibfield  {author} {\bibinfo {author} {\bibfnamefont {I.~I.}\ \bibnamefont
  {Yusipov}}, \bibinfo {author} {\bibfnamefont {O.~S.}\ \bibnamefont
  {Vershinina}}, \bibinfo {author} {\bibfnamefont {S.}~\bibnamefont {Denisov}},
  \bibinfo {author} {\bibfnamefont {S.~P.}\ \bibnamefont {Kuznetsov}}, \ and\
  \bibinfo {author} {\bibfnamefont {M.~V.}\ \bibnamefont {Ivanchenko}},\ }\href
  {\doibase 10.1063/1.5094324} {\bibfield  {journal} {\bibinfo  {journal}
  {Chaos}\ }\textbf {\bibinfo {volume} {29}},\ \bibinfo {pages} {063130}
  (\bibinfo {year} {2019})}\BibitemShut {NoStop}%
\bibitem [{\citenamefont {Ghose}\ \emph {et~al.}(2004)\citenamefont {Ghose},
  \citenamefont {Alsing}, \citenamefont {Deutsch}, \citenamefont
  {Bhattacharya},\ and\ \citenamefont
  {Habib}}]{ghoseTransitionClassicalChaos2004}%
  \BibitemOpen
  \bibfield  {author} {\bibinfo {author} {\bibfnamefont {S.}~\bibnamefont
  {Ghose}}, \bibinfo {author} {\bibfnamefont {P.}~\bibnamefont {Alsing}},
  \bibinfo {author} {\bibfnamefont {I.}~\bibnamefont {Deutsch}}, \bibinfo
  {author} {\bibfnamefont {T.}~\bibnamefont {Bhattacharya}}, \ and\ \bibinfo
  {author} {\bibfnamefont {S.}~\bibnamefont {Habib}},\ }\href {\doibase
  10/bt3m9g} {\bibfield  {journal} {\bibinfo  {journal} {Phys. Rev. A}\
  }\textbf {\bibinfo {volume} {69}},\ \bibinfo {pages} {052116} (\bibinfo
  {year} {2004})}\BibitemShut {NoStop}%
\bibitem [{\citenamefont {{Kumari,
  Meenu}}(2019)}]{kumarimeenuQuantumClassicalCorrespondenceEntanglement2019}%
  \BibitemOpen
  \bibfield  {author} {\bibinfo {author} {\bibnamefont {{Kumari, Meenu}}},\
  }\href {http://hdl.handle.net/10012/14860} {\emph {\bibinfo {title}
  {Quantum-{{Classical Correspondence}} and {{Entanglement}} in {{Periodically
  Driven Spin Systems}}}}}\ (\bibinfo  {publisher} {{UWSpace}},\ \bibinfo
  {year} {2019})\BibitemShut {NoStop}%
\bibitem [{\citenamefont {Pokharel}\ \emph {et~al.}(2018)\citenamefont
  {Pokharel}, \citenamefont {Misplon}, \citenamefont {Lynn}, \citenamefont
  {Duggins}, \citenamefont {Hallman}, \citenamefont {Anderson}, \citenamefont
  {Kapulkin},\ and\ \citenamefont
  {Pattanayak}}]{pokharelChaosDynamicalComplexity2018}%
  \BibitemOpen
  \bibfield  {author} {\bibinfo {author} {\bibfnamefont {B.}~\bibnamefont
  {Pokharel}}, \bibinfo {author} {\bibfnamefont {M.~Z.~R.}\ \bibnamefont
  {Misplon}}, \bibinfo {author} {\bibfnamefont {W.}~\bibnamefont {Lynn}},
  \bibinfo {author} {\bibfnamefont {P.}~\bibnamefont {Duggins}}, \bibinfo
  {author} {\bibfnamefont {K.}~\bibnamefont {Hallman}}, \bibinfo {author}
  {\bibfnamefont {D.}~\bibnamefont {Anderson}}, \bibinfo {author}
  {\bibfnamefont {A.}~\bibnamefont {Kapulkin}}, \ and\ \bibinfo {author}
  {\bibfnamefont {A.~K.}\ \bibnamefont {Pattanayak}},\ }\href {\doibase
  10.1038/s41598-018-20507-w} {\bibfield  {journal} {\bibinfo  {journal} {Sci.
  Rep.}\ }\textbf {\bibinfo {volume} {8}},\ \bibinfo {pages} {1} (\bibinfo
  {year} {2018})}\BibitemShut {NoStop}%
\bibitem [{\citenamefont {Ralph}\ \emph {et~al.}(2017)\citenamefont {Ralph},
  \citenamefont {Maskell},\ and\ \citenamefont
  {Jacobs}}]{ralphMultiparameterEstimationQuantum2017}%
  \BibitemOpen
  \bibfield  {author} {\bibinfo {author} {\bibfnamefont {J.~F.}\ \bibnamefont
  {Ralph}}, \bibinfo {author} {\bibfnamefont {S.}~\bibnamefont {Maskell}}, \
  and\ \bibinfo {author} {\bibfnamefont {K.}~\bibnamefont {Jacobs}},\ }\href
  {\doibase 10.1103/PhysRevA.96.052306} {\bibfield  {journal} {\bibinfo
  {journal} {Phys. Rev. A}\ }\textbf {\bibinfo {volume} {96}},\ \bibinfo
  {pages} {052306} (\bibinfo {year} {2017})}\BibitemShut {NoStop}%
\bibitem [{\citenamefont {Ott}\ \emph {et~al.}(1984)\citenamefont {Ott},
  \citenamefont {Antonsen},\ and\ \citenamefont
  {Hanson}}]{ottEffectNoiseTimeDependent1984}%
  \BibitemOpen
  \bibfield  {author} {\bibinfo {author} {\bibfnamefont {E.}~\bibnamefont
  {Ott}}, \bibinfo {author} {\bibfnamefont {T.~M.}\ \bibnamefont {Antonsen}}, \
  and\ \bibinfo {author} {\bibfnamefont {J.~D.}\ \bibnamefont {Hanson}},\
  }\href {\doibase 10.1103/PhysRevLett.53.2187} {\bibfield  {journal} {\bibinfo
   {journal} {Phys. Rev. Lett.}\ }\textbf {\bibinfo {volume} {53}},\ \bibinfo
  {pages} {2187} (\bibinfo {year} {1984})}\BibitemShut {NoStop}%
\bibitem [{\citenamefont {Dittrich}\ and\ \citenamefont
  {Graham}(1987)}]{dittrichQuantumEffectsSteady1987}%
  \BibitemOpen
  \bibfield  {author} {\bibinfo {author} {\bibfnamefont {T.}~\bibnamefont
  {Dittrich}}\ and\ \bibinfo {author} {\bibfnamefont {R.}~\bibnamefont
  {Graham}},\ }\href {\doibase 10.1209/0295-5075/4/3/002} {\bibfield  {journal}
  {\bibinfo  {journal} {Europhys. Lett.}\ }\textbf {\bibinfo {volume} {4}},\
  (\bibinfo {year} {1987})}\BibitemShut {NoStop}%
\bibitem [{\citenamefont {Klappauf}\ \emph {et~al.}(1998)\citenamefont
  {Klappauf}, \citenamefont {Oskay}, \citenamefont {Steck},\ and\ \citenamefont
  {Raizen}}]{klappaufObservationNoiseDissipation1998}%
  \BibitemOpen
  \bibfield  {author} {\bibinfo {author} {\bibfnamefont {B.}~\bibnamefont
  {Klappauf}}, \bibinfo {author} {\bibfnamefont {W.}~\bibnamefont {Oskay}},
  \bibinfo {author} {\bibfnamefont {D.}~\bibnamefont {Steck}}, \ and\ \bibinfo
  {author} {\bibfnamefont {M.}~\bibnamefont {Raizen}},\ }\href {\doibase
  10.1103/PhysRevLett.81.1203} {\bibfield  {journal} {\bibinfo  {journal}
  {Phys. Rev. Lett.}\ }\textbf {\bibinfo {volume} {81}},\  (\bibinfo {year}
  {1998})}\BibitemShut {NoStop}%
\bibitem [{\citenamefont {Klappauf}\ \emph {et~al.}(1999)\citenamefont
  {Klappauf}, \citenamefont {Oskay}, \citenamefont {Steck},\ and\ \citenamefont
  {Raizen}}]{klappaufErratumObservationNoise1999}%
  \BibitemOpen
  \bibfield  {author} {\bibinfo {author} {\bibfnamefont {B.}~\bibnamefont
  {Klappauf}}, \bibinfo {author} {\bibfnamefont {W.}~\bibnamefont {Oskay}},
  \bibinfo {author} {\bibfnamefont {D.}~\bibnamefont {Steck}}, \ and\ \bibinfo
  {author} {\bibfnamefont {M.}~\bibnamefont {Raizen}},\ }\href {\doibase
  10.1103/PhysRevLett.82.241} {\bibfield  {journal} {\bibinfo  {journal} {Phys.
  Rev. Lett.}\ }\textbf {\bibinfo {volume} {82}},\  (\bibinfo {year}
  {1999})}\BibitemShut {NoStop}%
\bibitem [{\citenamefont {Ammann}\ \emph {et~al.}(1998)\citenamefont {Ammann},
  \citenamefont {Gray}, \citenamefont {Shvarchuck},\ and\ \citenamefont
  {Christensen}}]{ammannQuantumDeltaKickedRotor1998}%
  \BibitemOpen
  \bibfield  {author} {\bibinfo {author} {\bibfnamefont {H.}~\bibnamefont
  {Ammann}}, \bibinfo {author} {\bibfnamefont {R.}~\bibnamefont {Gray}},
  \bibinfo {author} {\bibfnamefont {I.}~\bibnamefont {Shvarchuck}}, \ and\
  \bibinfo {author} {\bibfnamefont {N.}~\bibnamefont {Christensen}},\ }\href
  {\doibase 10.1103/PhysRevLett.80.4111} {\bibfield  {journal} {\bibinfo
  {journal} {Phys. Rev. Lett.}\ }\textbf {\bibinfo {volume} {80}},\  (\bibinfo
  {year} {1998})}\BibitemShut {NoStop}%
\bibitem [{\citenamefont {Pattanayak}\ \emph {et~al.}(2003)\citenamefont
  {Pattanayak}, \citenamefont {Sundaram},\ and\ \citenamefont
  {Greenbaum}}]{pattanayakParameterScalingDecoherent2003}%
  \BibitemOpen
  \bibfield  {author} {\bibinfo {author} {\bibfnamefont {A.~K.}\ \bibnamefont
  {Pattanayak}}, \bibinfo {author} {\bibfnamefont {B.}~\bibnamefont
  {Sundaram}}, \ and\ \bibinfo {author} {\bibfnamefont {B.~D.}\ \bibnamefont
  {Greenbaum}},\ }\href
  {https://link.aps.org/doi/10.1103/PhysRevLett.90.014103} {\bibfield
  {journal} {\bibinfo  {journal} {Phys. Rev. Lett.}\ }\textbf {\bibinfo
  {volume} {90}} (\bibinfo {year} {2003})}\BibitemShut {NoStop}%
\bibitem [{\citenamefont {Gong}\ and\ \citenamefont
  {Brumer}(2005)}]{gongQuantumChaosMeets2005}%
  \BibitemOpen
  \bibfield  {author} {\bibinfo {author} {\bibfnamefont {J.}~\bibnamefont
  {Gong}}\ and\ \bibinfo {author} {\bibfnamefont {P.}~\bibnamefont {Brumer}},\
  }\href {https://doi.org/10.1146/annurev.physchem.56.092503.141319} {\bibfield
   {journal} {\bibinfo  {journal} {Annu. Rev. Phys. Chem.}\ }\textbf {\bibinfo
  {volume} {56}} (\bibinfo {year} {2005})}\BibitemShut {NoStop}%
\bibitem [{\citenamefont {Habib}\ \emph {et~al.}(1998)\citenamefont {Habib},
  \citenamefont {Shizume},\ and\ \citenamefont
  {Zurek}}]{habibDecoherenceChaosCorrespondence1998}%
  \BibitemOpen
  \bibfield  {author} {\bibinfo {author} {\bibfnamefont {S.}~\bibnamefont
  {Habib}}, \bibinfo {author} {\bibfnamefont {K.}~\bibnamefont {Shizume}}, \
  and\ \bibinfo {author} {\bibfnamefont {W.~H.}\ \bibnamefont {Zurek}},\ }\href
  {\doibase 10.1103/PhysRevLett.80.4361} {\bibfield  {journal} {\bibinfo
  {journal} {Phys. Rev. Lett.}\ }\textbf {\bibinfo {volume} {80}},\ \bibinfo
  {pages} {4361} (\bibinfo {year} {1998})}\BibitemShut {NoStop}%
\bibitem [{\citenamefont {Percival}(1998)}]{percivalQuantumStateDiffusion1998}%
  \BibitemOpen
  \bibfield  {author} {\bibinfo {author} {\bibfnamefont {I.}~\bibnamefont
  {Percival}},\ }\href@noop {} {\emph {\bibinfo {title} {Quantum State
  Diffusion}}}\ (\bibinfo  {publisher} {{Cambridge University Press}},\
  \bibinfo {address} {{Cambridge, UK : New York}},\ \bibinfo {year}
  {1998})\BibitemShut {NoStop}%
\bibitem [{\citenamefont {Brun}\ \emph {et~al.}(1996)\citenamefont {Brun},
  \citenamefont {Percival},\ and\ \citenamefont
  {Schack}}]{brunQuantumChaosOpen1996}%
  \BibitemOpen
  \bibfield  {author} {\bibinfo {author} {\bibfnamefont {T.}~\bibnamefont
  {Brun}}, \bibinfo {author} {\bibfnamefont {I.}~\bibnamefont {Percival}}, \
  and\ \bibinfo {author} {\bibfnamefont {R.}~\bibnamefont {Schack}},\ }\href
  {\doibase 10.1088/0305-4470/29/9/020} {\bibfield  {journal} {\bibinfo
  {journal} {J. Phys. A-Math. Gen.}\ }\textbf {\bibinfo {volume} {29}},\
  (\bibinfo {year} {1996})}\BibitemShut {NoStop}%
\bibitem [{\citenamefont {Boffetta}\ \emph {et~al.}(2002)\citenamefont
  {Boffetta}, \citenamefont {Cencini}, \citenamefont {Falcioni},\ and\
  \citenamefont {Vulpiani}}]{boffettaPredictabilityWayCharacterize2002}%
  \BibitemOpen
  \bibfield  {author} {\bibinfo {author} {\bibfnamefont {G.}~\bibnamefont
  {Boffetta}}, \bibinfo {author} {\bibfnamefont {M.}~\bibnamefont {Cencini}},
  \bibinfo {author} {\bibfnamefont {M.}~\bibnamefont {Falcioni}}, \ and\
  \bibinfo {author} {\bibfnamefont {A.}~\bibnamefont {Vulpiani}},\ }\href
  {\doibase 10.1016/S0370-1573(01)00025-4} {\bibfield  {journal} {\bibinfo
  {journal} {Phys. Rep.}\ }\textbf {\bibinfo {volume} {356}},\  (\bibinfo
  {year} {2002})}\BibitemShut {NoStop}%
\bibitem [{\citenamefont {Wolf}\ \emph {et~al.}(1985)\citenamefont {Wolf},
  \citenamefont {Swift}, \citenamefont {Swinney},\ and\ \citenamefont
  {Vastano}}]{wolfDeterminingLyapunovExponents1985}%
  \BibitemOpen
  \bibfield  {author} {\bibinfo {author} {\bibfnamefont {A.}~\bibnamefont
  {Wolf}}, \bibinfo {author} {\bibfnamefont {J.}~\bibnamefont {Swift}},
  \bibinfo {author} {\bibfnamefont {H.}~\bibnamefont {Swinney}}, \ and\
  \bibinfo {author} {\bibfnamefont {J.}~\bibnamefont {Vastano}},\ }\href
  {\doibase 10.1016/0167-2789(85)90011-9} {\bibfield  {journal} {\bibinfo
  {journal} {Physica D}\ }\textbf {\bibinfo {volume} {16}},\  (\bibinfo {year}
  {1985})}\BibitemShut {NoStop}%
\bibitem [{\citenamefont {Berman}\ and\ \citenamefont
  {Zaslavsky}(1978)}]{bermanConditionStochasticityQuantum1978a}%
  \BibitemOpen
  \bibfield  {author} {\bibinfo {author} {\bibfnamefont {G.~P.}\ \bibnamefont
  {Berman}}\ and\ \bibinfo {author} {\bibfnamefont {G.~M.}\ \bibnamefont
  {Zaslavsky}},\ }\href {\doibase 10.1016/0378-4371(78)90190-5} {\bibfield
  {journal} {\bibinfo  {journal} {Physica A}\ }\textbf {\bibinfo {volume}
  {91}},\ \bibinfo {pages} {450} (\bibinfo {year} {1978})}\BibitemShut
  {NoStop}%
\bibitem [{\citenamefont
  {Prezhdo}(2002)}]{prezhdoClassicalMappingSecondorder2002}%
  \BibitemOpen
  \bibfield  {author} {\bibinfo {author} {\bibfnamefont {O.~V.}\ \bibnamefont
  {Prezhdo}},\ }\href {\doibase 10.1063/1.1493776} {\bibfield  {journal}
  {\bibinfo  {journal} {J. Chem. Phys.}\ }\textbf {\bibinfo {volume} {117}},\
  \bibinfo {pages} {2995} (\bibinfo {year} {2002})}\BibitemShut {NoStop}%
\bibitem [{\citenamefont {Tomsovic}\ and\ \citenamefont
  {Heller}(1991)}]{tomsovicSemiclassicalDynamicsChaotic1991}%
  \BibitemOpen
  \bibfield  {author} {\bibinfo {author} {\bibfnamefont {S.}~\bibnamefont
  {Tomsovic}}\ and\ \bibinfo {author} {\bibfnamefont {E.~J.}\ \bibnamefont
  {Heller}},\ }\href {\doibase 10.1103/PhysRevLett.67.664} {\bibfield
  {journal} {\bibinfo  {journal} {Phys. Rev. Lett.}\ }\textbf {\bibinfo
  {volume} {67}},\ \bibinfo {pages} {664} (\bibinfo {year} {1991})}\BibitemShut
  {NoStop}%
\bibitem [{\citenamefont {Banerjee}\ \emph {et~al.}(2004)\citenamefont
  {Banerjee}, \citenamefont {Bag}, \citenamefont {Banik},\ and\ \citenamefont
  {Ray}}]{banerjeeSolutionQuantumLangevin2004}%
  \BibitemOpen
  \bibfield  {author} {\bibinfo {author} {\bibfnamefont {D.}~\bibnamefont
  {Banerjee}}, \bibinfo {author} {\bibfnamefont {B.~C.}\ \bibnamefont {Bag}},
  \bibinfo {author} {\bibfnamefont {S.~K.}\ \bibnamefont {Banik}}, \ and\
  \bibinfo {author} {\bibfnamefont {D.~S.}\ \bibnamefont {Ray}},\ }\href
  {\doibase 10.1063/1.1711593} {\bibfield  {journal} {\bibinfo  {journal} {The
  Journal of Chemical Physics}\ }\textbf {\bibinfo {volume} {120}},\ \bibinfo
  {pages} {8960} (\bibinfo {year} {2004})}\BibitemShut {NoStop}%
\bibitem [{\citenamefont {Pattanayak}\ and\ \citenamefont
  {Schieve}(1994)}]{pattanayakGaussianWavepacketDynamics1994}%
  \BibitemOpen
  \bibfield  {author} {\bibinfo {author} {\bibfnamefont {A.}~\bibnamefont
  {Pattanayak}}\ and\ \bibinfo {author} {\bibfnamefont {W.}~\bibnamefont
  {Schieve}},\ }\href {\doibase 10.1103/PhysRevE.50.3601} {\bibfield  {journal}
  {\bibinfo  {journal} {Phys. Rev. E}\ }\textbf {\bibinfo {volume} {50}},\
  (\bibinfo {year} {1994})}\BibitemShut {NoStop}%
\bibitem [{\citenamefont {Misplon}\ and\ \citenamefont
  {Pattanayak}(2016)}]{misplon_unpublished}%
  \BibitemOpen
  \bibfield  {author} {\bibinfo {author} {\bibfnamefont {M.~Z.~R.}\
  \bibnamefont {Misplon}}\ and\ \bibinfo {author} {\bibfnamefont {A.~K.}\
  \bibnamefont {Pattanayak}},\ }\href@noop {} {\enquote {\bibinfo {title}
  {Investigation of co-existing attractors in the duffing oscillator},}\ }
  (\bibinfo {year} {2016}),\ \bibinfo {note} {unpublished}\BibitemShut
  {NoStop}%
\end{thebibliography}%
\bibliographystyle{apsrev4-1}

\clearpage
\appendix

\section{Derivation of the four equation semiclassical model}\label{sec:appendix}
The dissipative dynamics of the open semiclassical Duffing oscillator results in a conserved quantity, corresponding to reducing the wavefunction to one which is a minimum
uncertainty wavepacket. This enables the five equation semiclassical Duffing oscillator model from \cite{pokharelChaosDynamicalComplexity2018} to be reduced to four equations. 
To be more explicit, 
\begin{widetext}
\begin{equation}
dx = pdt + 2 \sqrt{\Gamma}((\mu - \frac{1}{2}) d\xi_R - Rd\xi_I),
\label{eq:SCdXOriginal}
\end{equation}
\begin{equation}
\begin{split}
    dp =  (-\beta^2(x^3 + 3 \mu x) + x - 2\Gamma_p + \frac{g}{\beta} \cos\omega t) dt + 2 \sqrt{\Gamma}(Rd\xi_R - (\kappa - \frac{1}{2}) d\xi_I),
\end{split}
\label{eq:SCdPOriginal}
\end{equation}
\begin{equation}
\begin{split}
\frac{d\mu}{dt} = 2R + 2\Gamma(\mu - \mu^2 - R^2 + \frac{1}{4}),
\label{eq:SCdMu}
\end{split}
\end{equation}
\begin{equation}
\begin{split}
\frac{d\kappa}{dt} = 2R(-3\beta^2x^2+1) + 2\Gamma(-\kappa - \kappa^2 - R^2 + \frac{1}{4}),
\label{eq:SCdKappa}
\end{split}
\end{equation}
\begin{equation}
\begin{split}
\frac{dR}{dt} = \mu(-3\beta^2x^2+1) + \kappa - 2\Gamma R(\mu + \kappa),
\label{eq:SCdR}
\end{split}
\end{equation}
\end{widetext}
can be reduced to \cref{eq:SCdx,eq:SCdp,eq:SCdrho,eq:SCdpi}.  The minimum uncertainty condition that allows for this reduction is
\begin{equation} \label{eq:minUncertaintyCondition}
\mu\kappa - R^2 = \frac{1}{4}.
\end{equation} 
The time derivative of $\mu\kappa - R^2$ can be shown to be of the form $\dot{X} = -2 \Gamma(\mu + \kappa) X$ and this guarantees the minimum uncertainty condition is quickly met when \(2\Gamma(\mu + \kappa)>0\) and \(t\gg1/2\Gamma(\mu + \kappa)\). We empirically confirm that the convergence is rapid. We then eliminate \(\kappa\) by
\begin{equation}
    \kappa = \frac{R^2  +\frac{1}{4}}{\mu}.
    \label{eq:kappa}
\end{equation}
We make two changes of variables: $\mu = \rho^2$ and $R = \rho \Pi$ \cite{pattanayakGaussianWavepacketDynamics1994}. It can be shown with \cref{eq:kappa} that \cref{eq:SCdMu} becomes
\begin{equation}
 \frac{d\rho}{dt} = \Pi + \Gamma \bigg( \rho - \rho^3 - \rho\Pi^2 + \frac{1}{4\rho} \bigg),
\label{eq:SCdrho2}
\end{equation}
which matches \cref{eq:SCdrho}. We use \(R = \rho \Pi\) and \cref{eq:SCdR} to show that
\begin{equation}
\begin{split}
  \frac{dR}{dt} = \Pi\frac{d\rho}{dt} + \rho\frac{d\Pi}{dt} = \mu(-3\beta^2x^2+1) + \kappa -
  2\Gamma R(\mu + \kappa).
\end{split}
\end{equation}
Using \(d\rho/dt\) and \(\kappa\), we determine
\begin{equation}
\begin{split}
  \frac{d\Pi}{dt} = \rho(-3\beta^2x^2+1) + \frac{1}{4\rho^3} - \Gamma\Pi \bigg( 1 + \Pi^2 +
  \rho^2 + \frac{3}{4\rho^2} \bigg),
\end{split}
\end{equation}
which matches \cref{eq:SCdpi}. Note that in the absence of environmental coupling (\(\Gamma = 0 \)), this system corresponds to an $x$-oscillator coupled to an $\rho$-oscillator \cite{pattanayakGaussianWavepacketDynamics1994}
\begin{equation}
  U(x,\rho) = -\frac{x^2}{2} + \beta^2(\frac{x^2}{2})^2 - (\frac{g}{\beta}\cos\omega t)x \\ - \frac{\rho^2}{2} + \frac{1}{8\rho^2} + \frac{3}{2}\beta^2(x\rho)^2.
\end{equation}

  \section{Causes of divergence in $d$}\label{sec:d_divergence}

  \begin{figure}
    \centering
    \includegraphics[width=\linewidth]{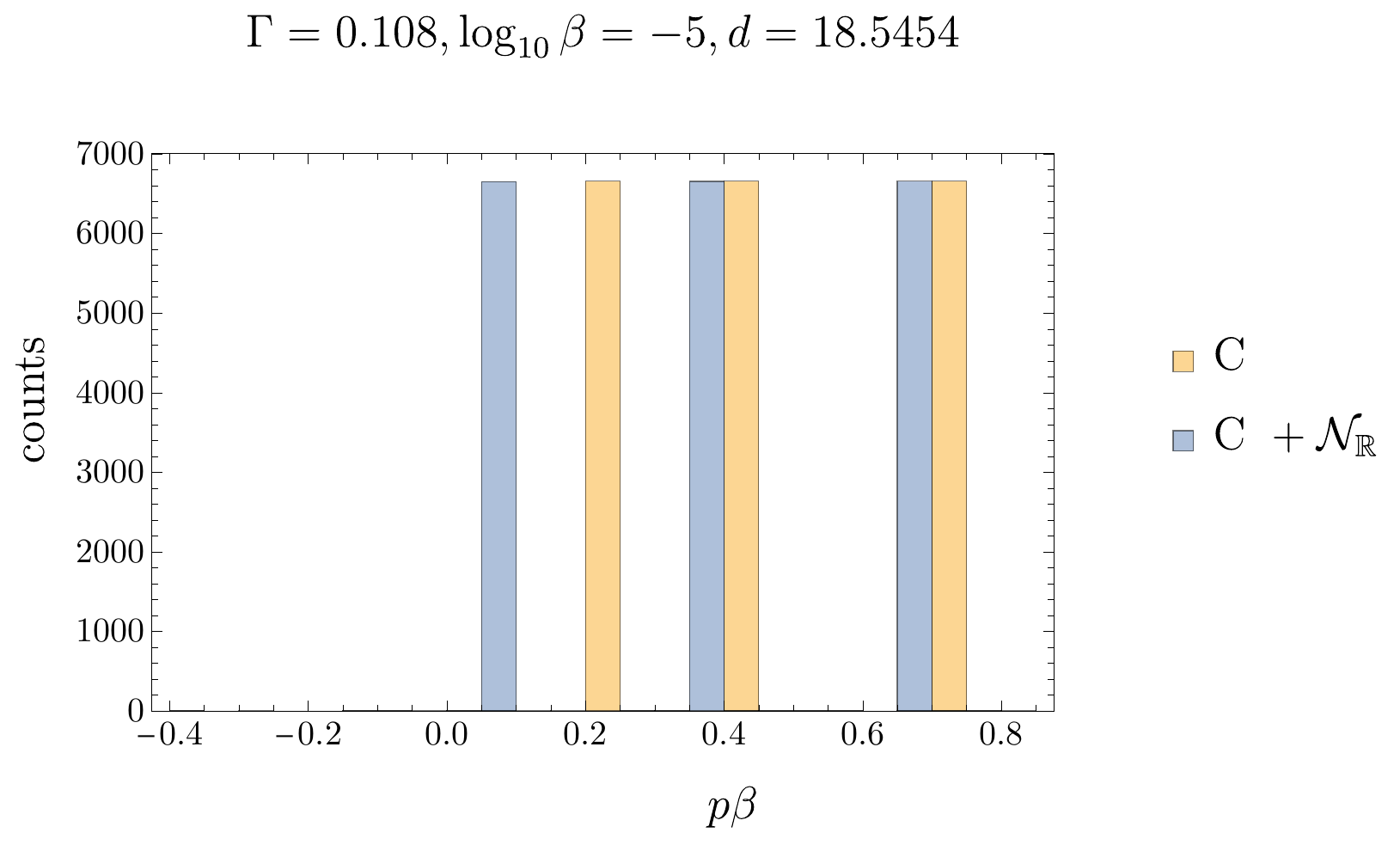}
    \caption{Although the \(\text{C}\) and \(\text{C}+\mathcal{N}_\mathbb{R}\) models have almost identical lyapunov exponent and similar spatial behavior as apparent in this plot, the distance measure $d$ fails to capture the similarities between distributions where the peaks in the distributions are offset by small amounts.}
    \label{fig:histogramchaos}
\end{figure}

 \begin{figure}
    \centering
    \includegraphics[width=\linewidth]{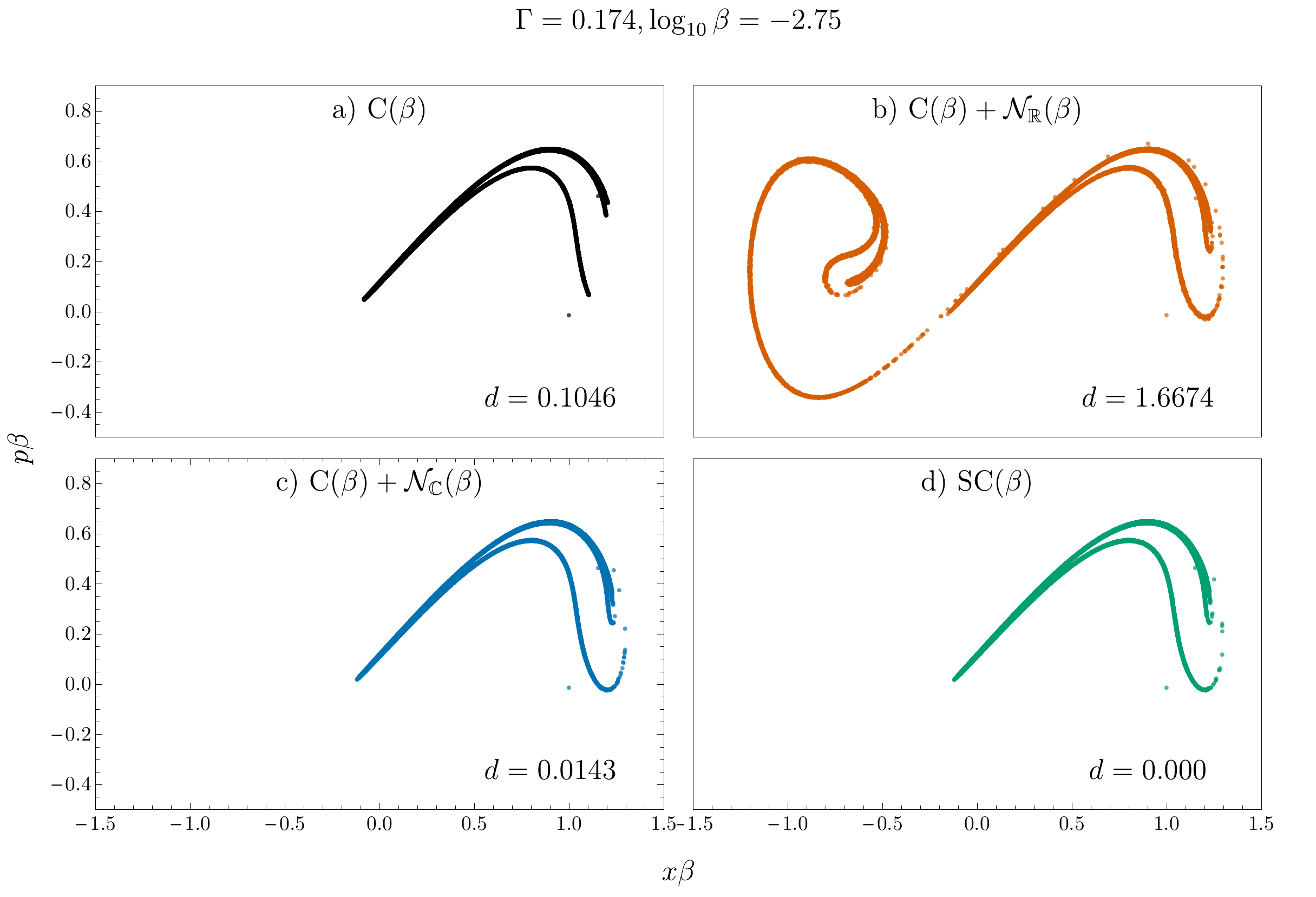}
    \caption{Different approximations are compared to the semiclassical model at the cusp of a transition from single to double well orbits at $\Gamma = 0.174$, $\log_{10}\beta = -2.75$.  
}
    \label{fig:poincarechaos}
\end{figure}

We note that the distance metric \(d\) can yield meaningless results when comparing periodic attractors. 
  This metric compares the distributions in \(x\) and \(p\), meaning that even slight displacements in sparse distributions of periodic attractors 
  will cause the metric to diverge. An example of this is shown in \cref{fig:histogramchaos}. The models \(\text{C}\) and
  \(\text{C}+\mathcal{N}_\mathbb{R}\) are quite similar and have nearly equal \(\lambda\), but the distance \(d=18.5454\) is really large. This pattern repeated in other
  examples of distances measured between periodic trajectories. For this reason, the lower
  bound of plots in \cref{fig:lyapKLgrid} is set at \(\log_{10}(\beta)=-3\), above the largest length scale at which this anomaly is present in the result.

  The distance metric also fails in the vicinity of discontinuous changes of spatial behavior. The anomalous spike in \(\Delta
\overline{d}_{\text{SC},\text{C}+\mathcal{N}_\mathbb{R}}\) at \(\log_{10} \beta = -2.75\) in the bottom left and right subplots of \cref{fig:lyapKLgrid} is caused by a single
attractor, \(\Gamma=0.198\), on the cusp of a transition from a single to double well orbit. This is evident in
\cref{fig:poincarechaos}, where the Poincare sections of C, \(\text{C}+\mathcal{N}_\mathbb{R}\), and SC show
trajectories constrained to the \(+x\) well while the Poincare section of \(\text{C}+\mathcal{N}_\mathbb{C}\) shows
trajectory in both the \(+x\) and \(-x\) wells. The \(\text{C}+\mathcal{N}_\mathbb{C}\) model jumps
the gun on the discontinuous change from single to double well trajectories, resulting an anomalously high
distance to SC.

\section{Energy Spectra Analysis}

  \begin{figure}[h!]
    \centering
    \includegraphics[width=\linewidth]{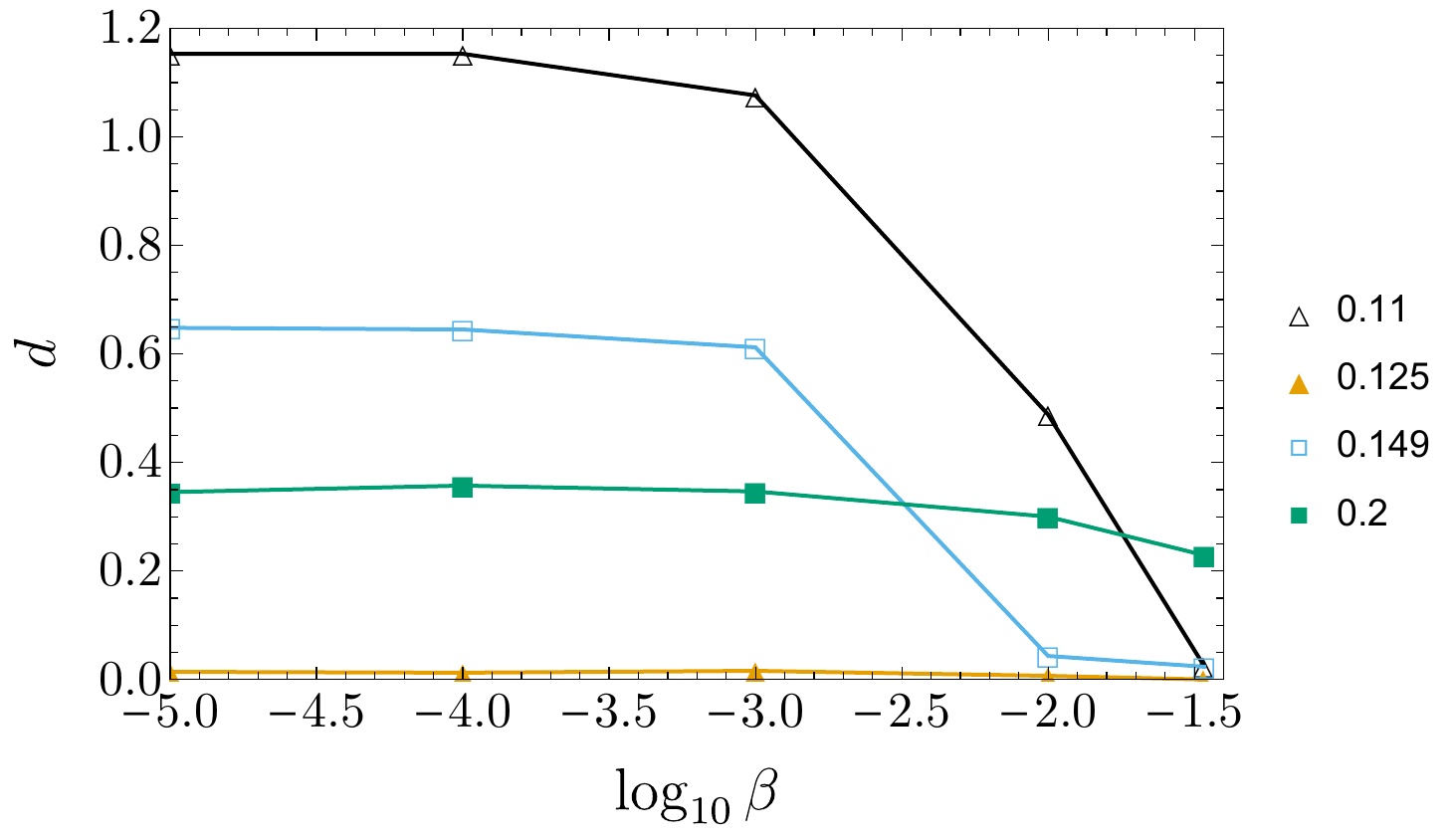}
    \caption{The distance between the energy spectra in \cref{fig:energy_histogram} and the spectra of \(\Gamma=0.125\) at \(\beta_{\text{conv}}\). The distance decreases as \(\beta \rightarrow \beta_{\text{conv}}\).}
    \label{fig:energy_distance}
  \end{figure}

\begin{figure*}[h!]
    \centering
    \includegraphics[width=\linewidth]{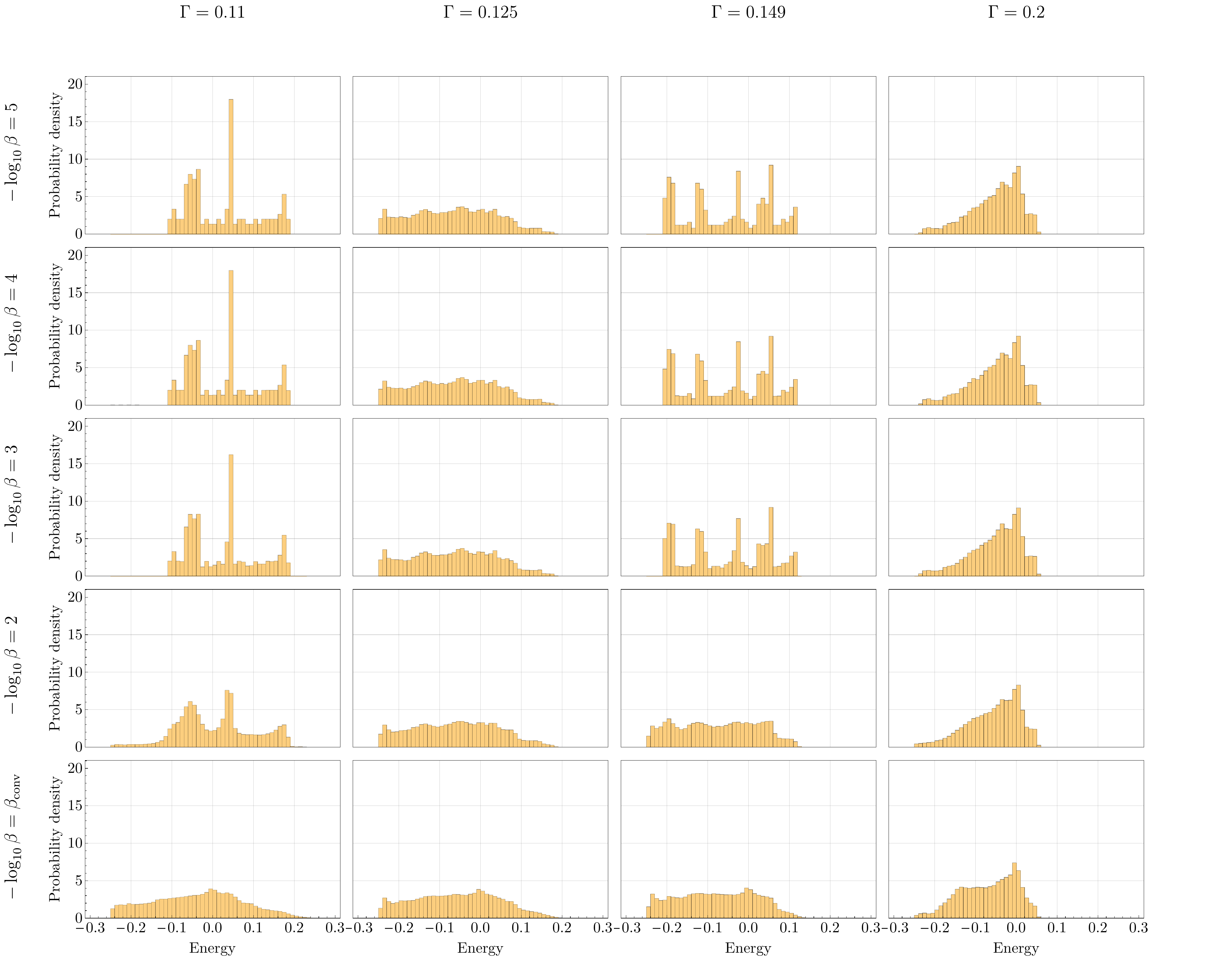}
    \caption{Energy spectra for attractors with coupling \(\Gamma = 0.11, 0.125, 0.149, 0.2\) for length scales between \(\beta =0.00001\) and \(\beta_{\text{conv}}=0.0341\). As \(\beta \rightarrow \beta_{\text{conv}}\), the spectra become more similar to one another.}
    \label{fig:energy_histogram}
  \end{figure*}

The convergence of attractors in the semiclassical regime is also evident in energy spectra.  \cref{fig:energy_histogram} shows energy spectra for attractors with coupling \(\Gamma = 0.11, 0.125, 0.149, 0.2\) for length scales between \(\beta =0.00001\) and \(\beta_{\text{conv}}=0.0341\). Note that attractors (\(\Gamma = 0.125, 0.2\)) are classically chaotic and
(\(\Gamma = 0.11, 0.149\)) are classically periodic. These attractors are generally representative of energy spectra in the intermediate coupling regime. The spectra are increasingly similar as \(\beta \rightarrow \beta_{\text{conv}}\), although \(\Gamma = 0.2\) at the edge of the intermediate coupling regime is not as converged as the others. We quantify this convergence by
measuring the distance \(d\) between the given spectra and the spectra of \(\Gamma=0.125\) at
\(\beta_{\text{conv}}\) (see \cref{fig:energy_distance}).  As expected, the distance decreases as \(\beta\) increases, illustrating a convergence of attractors.

\section{Classical noise-added models}

\begin{figure*}[h!]
    \centering
    \includegraphics[width=\linewidth]{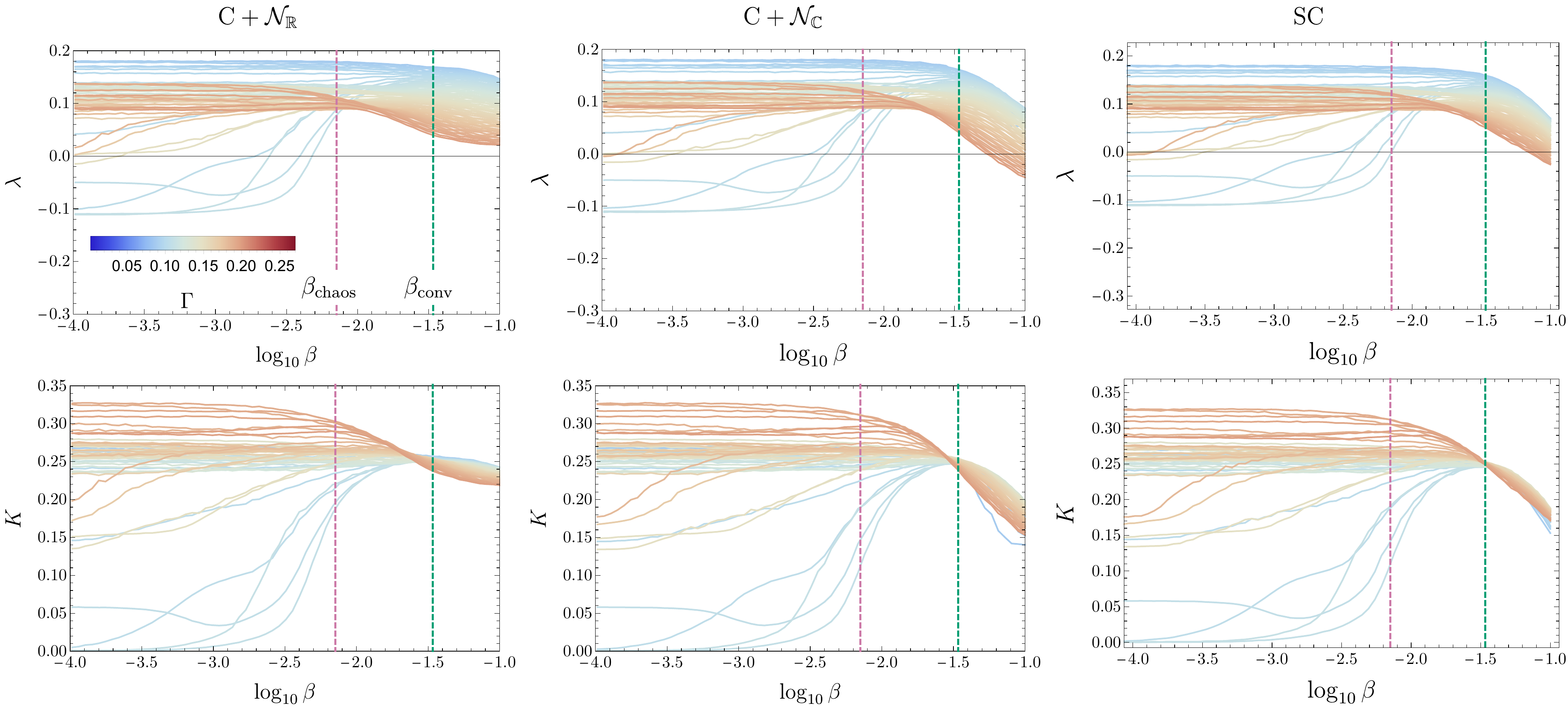}
    \caption{The \(\lambda(\beta;\Gamma)\) and \(K(\beta;\Gamma)\) curves for the two classical noise-added models (\(\text{C}+\mathcal{N}_\mathbb{R}\) and
\(\text{C}+\mathcal{N}_\mathbb{C}\)) and the semiclassical model (SC) over the intermediate damping regime (\(0.088 \leq \Gamma < 0.202\)). \(\Gamma\) is sampled at intervals of \(0.002\).}
    \label{fig:ClassicalNoiseK}
  \end{figure*}

\cref{fig:ClassicalNoiseK} shows \(K(\beta;\Gamma)\) for \(\text{C}+\mathcal{N}_\mathbb{R}\),
\(\text{C}+\mathcal{N}_\mathbb{C}\), and SC in the intermediate coupling regime. While the qualitative behavior of for the classical noise added models is similar to SC, there are noticeable
differences as highlighted in the main text. The variation between the models is more pronounced as \(\log_{10}\beta\) increases past \(\log_{10}\beta_{\text{conv}}\). We note that the \(K\) convergence is not unique to the SC model. The convergence in \(K\) can be caused by simply adding classical noise to the Duffing oscillator as well; it does not require a quantum description. 

\section{Duffing oscillator behavior in lower and higher environmental coupling regimes}\label{sec:low_and_high}

\begin{figure*}[h!]
    \centering
    \includegraphics[width=\linewidth]{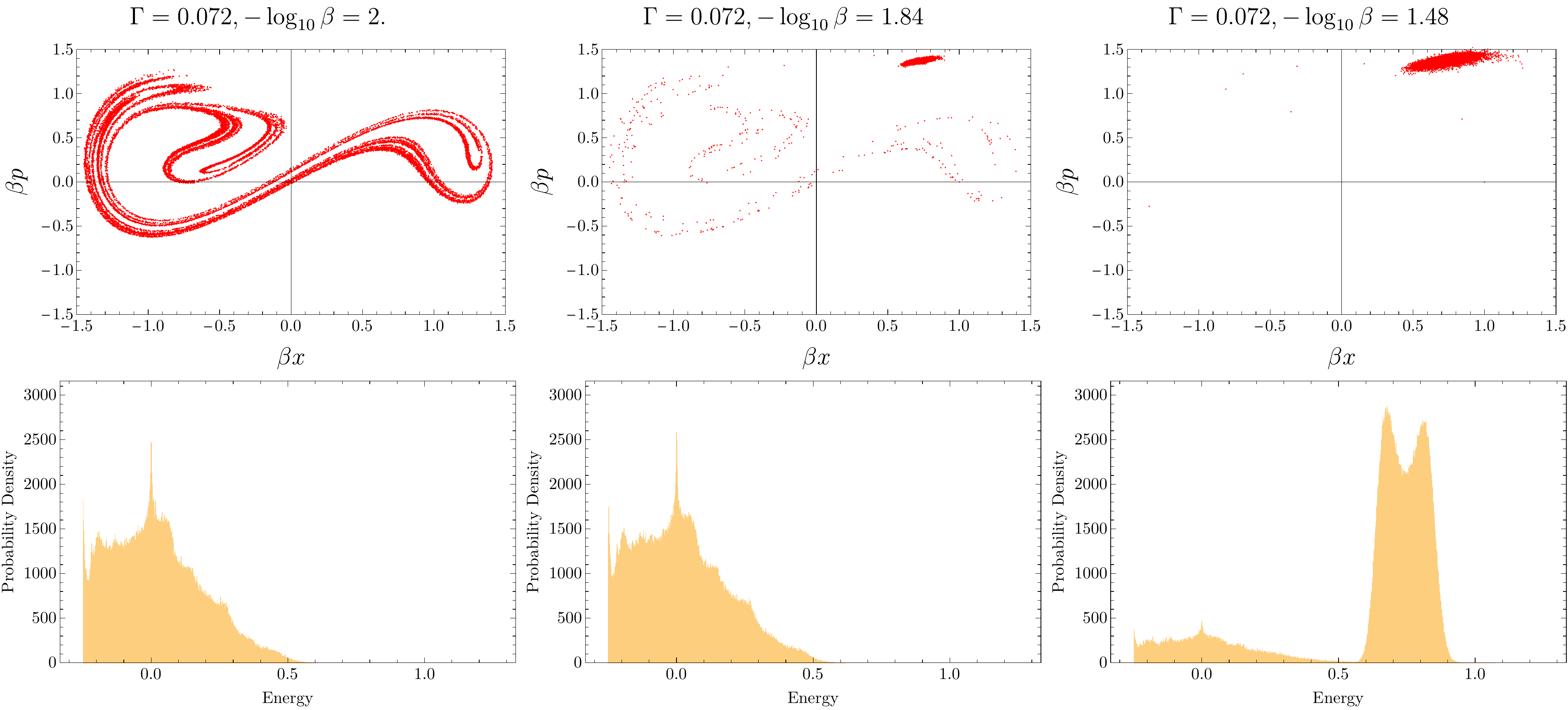}
    \caption{An example of the transition between chaotic and periodic motion for coupling levels between the low and intermediate range visualized with energy spectra and Poincare sections. Attractors in this range ($0.068 \leq \Gamma < 0.88$)
     all exhibit a similar transition.}
    \label{fig:CoexistingAttractors}
  \end{figure*}

\begin{figure*}[h!]
    \centering
    \includegraphics[width=\linewidth]{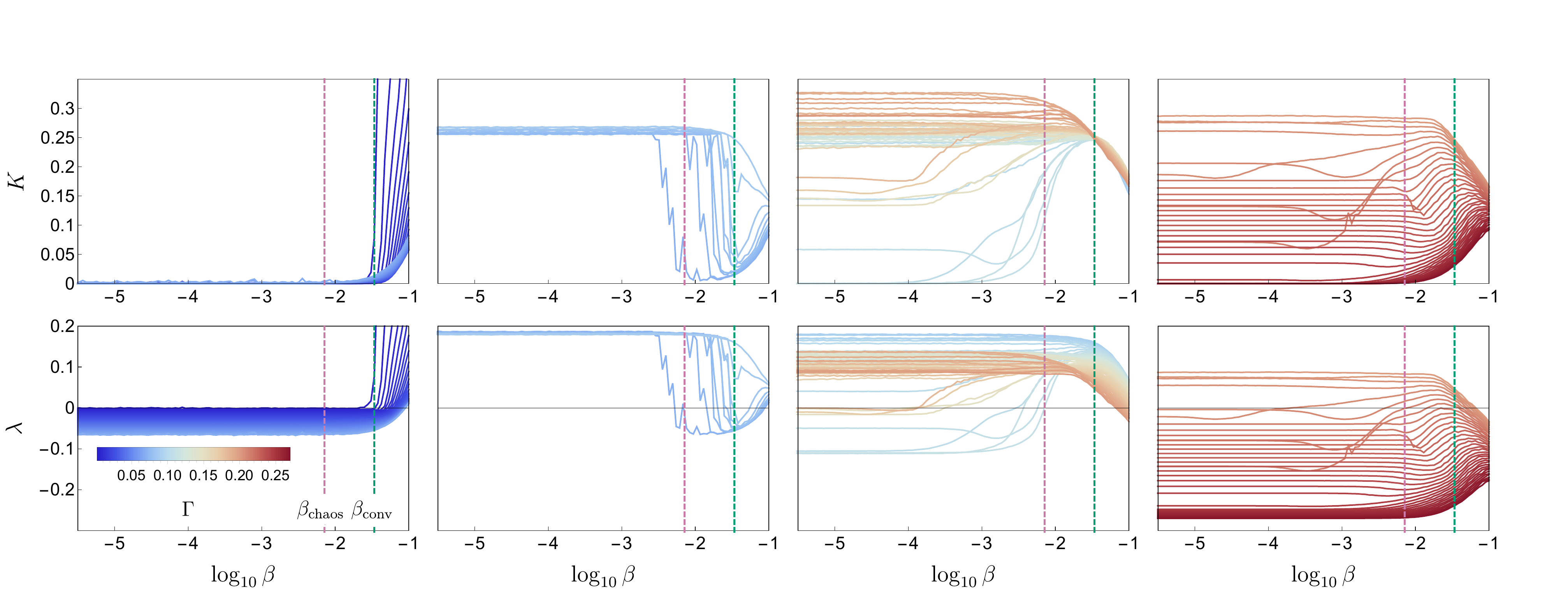}
    \caption{The top row of subplots show \(\lambda\) as a function of length scale (\(\log_{10}(\beta)\)) for the semiclassical model over four ranges of increasing \(\Gamma\)
      that correspond to distinct regions of behavior for \(\lambda\) and \(K\). The bottom
      row shows the same ranges of \(\Gamma\) values for \(K\) as a function of length scale,
      \(\log(\beta^{-1})\).}
    \label{fig:GridOfLambdaAndK}
\end{figure*}

We present our investigation of lower and higher coupling regimes here for the sake of completeness. In the low coupling regime, \(\Gamma < 0.068\), we observe
\(\lambda=-\Gamma\) in classical and most semiclassical length scales, as seen in \cref{fig:GridOfLambdaAndK}. The oscillator traverses both wells in single period orbits.  
At high \(\beta\), we observe a rise in \(\lambda\) that aligns with the `chaotification' of low coupling periodic orbits discovered earlier ~\cite{pokharelChaosDynamicalComplexity2018}. However, these values of small $\Gamma$ and large $\beta$ are beyond the range of validity of the semiclassical formalism, and we do not attribute any physical meaning to this rapid change in $\lambda$. 

Between the low and intermediate coupling regimes (\(0.068 \leq \Gamma < 0.088\)), classically chaotic orbits sharply transition to periodic behavior at some semiclassical length scale \(-3 < \log_{10} \beta < \log_{10} \beta_{\text{conv}}\) (see \cref{fig:GridOfLambdaAndK}). During the transition from chaotic to periodic behavior, the oscillator gradually localizes to a single well and a double-peaked, high-energy cluster emerges in the energy spectra (see \cref{fig:CoexistingAttractors}). We suspect that 1) the POs at high \(\beta\) in this coupling regime are linked to the classical POs in the low coupling regime and 2) the chaotic orbits at low \(\beta\) are linked to the classical chaotic orbits in the intermediate regime. Given that previous work has shown that the Lyapunov exponent is sensitive to initial conditions for some attractors in this regime \cite{misplon_unpublished}, further exploration is warranted. The overall effect of this phenomena is that increasing \(\beta\) results in a delay to the onset of chaos with respect to coupling \(\Gamma\) (see \cref{fig:lyap}).
In the high coupling regime (\(\Gamma > 0.2\)), \(\lambda\) falls off steeply with higher \(\Gamma\) (see \cref{fig:lyap}). Here the oscillator is increasingly limited to a single well because of heavy dissipation through the environmental coupling.  For \(\Gamma \geq 0.25\),
the oscillator obeys \(\lambda = - \Gamma\) for much of the semiclassical regime. Similar to the low coupling regime, \(\lambda\) monotonically rises at sufficiently large \(\beta\).

\end{document}